\pdfoutput=1

\documentclass[twocolumn]{aastex63}

\received{XXXX}
\revised{YYYY}
\accepted{ZZZZ}
\submitjournal{ApJ}

\shorttitle{Dust filaments in molecular cloud envelopes}
\shortauthors{Beitia-Antero et al.}

\begin{document}

\title{Formation of dust filaments in the diffuse envelopes of molecular clouds}

\correspondingauthor{Leire Beitia-Antero}
\email{lbeitia@ucm.es}

\author[0000-0003-0833-4075]{Leire Beitia-Antero}
\affiliation{Joint Center for Ultraviolet Astronomy \\
Universidad Complutense de Madrid \\
Edificio Fisac (Fac. Estudios Estad\'isticos), Av. Puerta de Hierro s/n, 28040, Madrid, Spain}
\affiliation{Departamento de F\'isica de la Tierra y Astrof\'isica (U.D. Astronom\'ia y Geodesia)\\
Facultad de CC. Matem\'aticas, Universidad Complutense de Madrid \\
Plaza de las Ciencias 3, 28040, Madrid, Spain}

\author{Ana I. G\'omez de Castro}
\affiliation{Joint Center for Ultraviolet Astronomy \\
Universidad Complutense de Madrid \\
Edificio Fisac (Fac. Estudios Estad\'isticos), Av. Puerta de Hierro s/n, 28040, Madrid, Spain}
\affiliation{Departamento de F\'isica de la Tierra y Astrof\'isica (U.D. Astronom\'ia y Geodesia)\\
Facultad de CC. Matem\'aticas, Universidad Complutense de Madrid \\
Plaza de las Ciencias 3, 28040, Madrid, Spain}

\author[0000-0002-7409-0868]{Juan C. Vallejo}
\affiliation{Joint Center for Ultraviolet Astronomy \\
Universidad Complutense de Madrid \\
Edificio Fisac (Fac. Estudios Estad\'isticos), Av. Puerta de Hierro s/n, 28040, Madrid, Spain}
\affiliation{Departamento de F\'isica de la Tierra y Astrof\'isica (U.D. Astronom\'ia y Geodesia)\\
Facultad de CC. Matem\'aticas, Universidad Complutense de Madrid \\
Plaza de las Ciencias 3, 28040, Madrid, Spain}



\begin{abstract}
  The path to understanding star formation processes begins
  with the study of the formation of molecular clouds. The
  outskirts of these clouds are characterized by
    low column densities that allow the penetration of
  ultraviolet radiation, resulting in a non-negligible ionization
fraction and the charging of the small dust grains that
are mixed with the gas; this diffuse phase is then coupled to
  the ambient magnetic field. \\
Despite the general assumption that dust and gas are
tightly correlated, several observational and theoretical
studies have reported variations in the dust-to-gas ratio toward
diffuse and cold clouds.  \\
In this work, we present the implementation of a new charged particles
module for analyzing the dust dynamics in molecular cloud
envelopes. We
study the evolution of a single population
of small charged grains (0.05 $\mu$m) in the turbulent, magnetized
molecular cloud envelope
using this module.
We show that variations in the dust-to-gas ratio arise due to
the coupling of the grains with the magnetic field, forming
elongated dust structures decoupled from the gas. This emphasizes
the importance of considering the dynamics of charged dust
when simulating the different phases of the interstellar medium, especially
for star formation studies.

\end{abstract}



\section{Introduction} \label{sec:intro}

Molecular clouds (MCs) are the galactic reservoirs of molecular material
where star formation takes place. It is therefore necessary
to understand their formation and evolution if
one aspires to comprehend the
physical processes involved in star formation. \par

It is known that MCs are formed from a more diffuse
medium that becomes thermally unstable
(see the recent reviews by \textit{e.g.}
\citealt{2020SSRv..216...76B} and \citealt{2020SSRv..216...50C}). Their
turbulent nature \citep{2004ARA&A..42..211E} is inherited from the parental gas
\citep{2018ApJ...854..100M,2019MNRAS.484.5734K}, and is
mantained by several processes both at large and small scales,
such as supernova explosions
\citep{2016ApJ...822...11P,2020arXiv200610764B}
or stellar jets
\citep{2015MNRAS.450.4035F,2018NatAs...2..896O}. This turbulence,
together with the pervasive
galactic magnetic field, plays a fundamental role in the
fragmentation of the cloud and the formation of dense gas filaments where
stars are actually formed
\citep{1988ApJ...329..392M,2000MNRAS.311..105F,2008ApJ...687..354N,2015A&A...580A..49I}.\par

MCs are formed from the accretion of warm, diffuse gas
from the warm neutral medium (WNM), 
characterized by its atomic composition (mainly HI),
temperatures of approximately 6000K,
densities ranging from 0.2 - 0.5 cm$^{-3}$ \citep{2001RvMP...73.1031F}
and ionization fractions of $\chi \sim 0.1$ \citep{2016SAAS...43...85K};
these conditions are sustained in time due to the
  constant bombardement of the interstellar radiation field over
the molecular gas.
Therefore, MC envelopes share properties with the WNM, and
they constitute the transition layers between the hot ionized gas of the
diffuse interstellar medium (ISM)
and the interiors of the clouds. The net charge of the
species in the outskirts of these
clouds favours their coupling
with the galactic magnetic field, which act as a carrier of
hydromagnetic waves \citep{1996ApJ...465..775B}; they
may also facilitate the penetration of
cosmic rays inside the cloud \citep{2018ApJ...855...23I}. \par

A key ingredient in the transition from atomic to molecular
gas is interstellar dust, which distribution
seems to be correlated with that of the gas
\citep{1978ApJ...224..132B,1996A&A...312..256B,2010A&A...518L..74R,2017ApJ...846...38L}.
Despite representing only 1\% in
mass of the ISM, interstellar dust grains play a fundamental
role in several physical phenomena. They act as catalysts in
chemical reactions
\citep{1971ApJ...163..155H,2004ApJ...604..222C,2020ApJ...888...52S},
shield the interior of the clouds from the
high energy ultraviolet (UV) radiation, and contribute
to the thermal regulation of the cloud
\citep{1994ApJ...427..822B,1995ApJ...443..152W,2001ApJS..134..263W,2010A&A...510A.110H} \\
Due to their composition,  dust grains acquire
a net charge in the presence of a radiation field
\citep{2001ApJS..134..263W}; even in the densest regions of MCs,
grains may be charged due to the action of
cosmic rays \citep{2015ApJ...812..135I}.
In MC envelopes, dust grains contribute to the
magnetic field coupling and affect
the propagation of hydromagnetic waves
\citep{1987ApJ...314..341P,1997PASA...14..170C} and, in last
instance, star formation \citep{1998ApJ...494..587N}. It is then
crucial to consider interstellar dust grains when
studying the formation and evolution of MCs.


Over the last decade, our knowledge of the cold
ISM 
has grown exponentially thanks to the results of
dedicated missions such as \textit{Herschel}
(see \textit{e.g.}
\citealt{2010A&A...518L.102A,2011A&A...529L...6A,2013A&A...550A..38P})
and \textit{Planck} \citep{2018arXiv180104945P,2018arXiv180706212P}.
However, the characterization of the diffuse ISM is more
challenging, since it requires tracers sensitive to low column
densities, such as $^{12}$CO
\citep{1996ApJ...472..191F,2005A&A...433..997F}
or very high-resolution HI maps
  \citep{2014ApJ...789...82C,2015ApJ...809..153M}.
An alternative approach, which is
very time-consuming but provides a higher resolution,
is to perform a spectroscopic
survey of the neutral and ionized components of the ISM
at optical and UV wavelengths
\citep{1991ApJ...379..245S,2003A&A...411..447L,2004AdSpR..34...41R,2010A&A...510A..54W}.

With these ingredients, it
seems that a full characterization of MCs may be within reach. However,
observations give at most a line-of-sight averaged view of the processes
involved
in MC evolution and thus, to understand the underlying physics, numerical
simulations of these processes are required. \par

It has been predicted that negatively charged grains may affect the
star formation efficiency due to gyroresonance effects with MHD waves
propagating inside MCs \citep{1987ApJ...314..341P,1998ApJ...494..587N,2003ApJ...597..970F,2020MNRAS.496.2123H}. For instance, \citet{2006MNRAS.371..513C}
demonstrated that C-type MHD shocks profiles are deeply affected
by the presence of a population of charged grains; \citet{2019AJ....157...83P}
have recently shown that the Kevin-Helmholz instability in MCs
depends on the dust charge density and size distribution; and
\citet{2007Ap&SS.311...35W} and \citet{2008A&A...484....1P} derived the modified coefficients for
non-ideal MHD effects (Ohmic resistivity, and ambipolar and Hall
diffusion) in the presence of charged dust grains. However,
none of the abovementioned studies followed the numerical evolution
of a MC in the presence of a population of dust.\par

The main trend has been
to consider the aerodynamic drag of dust in the framework of
protoplanetary disks \citep{2010ApJS..190..297B,2018PASA...35...31P,2019ApJS..244...38M}. Up to the knowledge of the authors, the only ones
that have included the physics of charged grains in their
MHD simulations are \citet{2017MNRAS.469.3532L}; they have
devoted a series of papers to the exhaustive
study of resonant instabilities that arise between
gas and charged dust over different phases of the ISM when
dust motions are driven by an external force
\citep{2018MNRAS.479.4681H,2019MNRAS.485.3991S,2020MNRAS.496.2123H}. \par

In this article, we study the dynamics of both gas and charged
dust under conditions
typical of a molecular cloud envelope
with a modified version
of the
MHD code Athena\footnote{\url{https://princetonuniversity.github.io/Athena-Cversion/}} \citep{2008ApJS..178..137S}.
In Section \ref{sec:eom_dust}
we present our implementation of the dynamics
of charged dust. In Section \ref{sec:simulations}, we present
the numerical 2D model for the MC envelope (\ref{subsec:numerical_model})
and analyze the local variations of the dust-to-gas ratio
(\ref{subsec:gas_dust}) as well as the dust filamentary
structure (\ref{subsec:dust_filaments}). In Section
\ref{sec:discussion}, a discussion of the results is
provided and the main results are highlighted
in Section \ref{sec:conclusions}.

\section{Numerical modeling of charged dust}\label{sec:eom_dust}
We worked with the latest stable version of 
Athena (v4.2). This C-version  has already implemented
an aerodynamic particles module (\citealt{2010ApJS..190..297B}, hereafter BS2010). In this section, we detail how we have expanded this module by adding
the necessary equations that take into account the 
dynamics of charged dust. We have
also carried out different tests for checking this new implementation
that are detailed in Appendix \ref{appendix:code_test}.\par

Apart from
the traditional drag term characterized by the stopping
time $t_{s}$ \citep{1977MNRAS.180...57W}, we added two terms to the equation
of motion of a charged particle that account for
the Coulomb drag from charged gas species, and the Lorentz
force:

\begin{equation}
  \frac{d {\bf v}}{d t} = - \frac{{\bf v} - {\bf u}}{t_{s}} -
  \nu_{C}({\bf v} - {\bf u})
  + \frac{Z_{d}e}{m_{d}c}({\bf v} - {\bf u})  \times {\bf B}
  \label{eq:dust_eom}
\end{equation}

In the above equation, {\bf v} is the velocity of the
dust grain, {\bf u} is the gas velocity, $t_{s}$ is the
aerodynamic stopping time, $Z_{d}$ is the
grain charge, $m_{d}$ is the dust mass, $c$ is the speed of light and
the term $\nu_{C}$ is an additional drag term that accounts for
Coulomb interactions between charged grains and ions/electrons
in the plasma:

\begin{equation}
  \nu_{C} =  \frac{4}{3}\frac{\sqrt{2\pi}Z_{d}^{2}e^{4}\ln \Lambda n_{d}}{(kT)^{3/2}}\bigg( \frac{\delta}{\sqrt{m_{ion}}} + \frac{(1-\delta)}{\sqrt{m_{e}}} \bigg)
  \label{eq:coulomb_rate}
\end{equation}

In this equation, $\ln \Lambda = \ln [(3kT/2e^{3})(kTm_{e}/\rho_{e}\pi)^{1/2}]$
is the Coulomb logarithm, $n_{d}$ is the
density of charged dust grains, and $\delta = 1$ if $Z < 0$,
$\delta = 0$ otherwise; this
means that negatively (positively) charged dust grains interact with
the ions (electrons) in the gas. We disregarded the effect
of the electric field derived from the magnetic field in the Lorentz
force because we are mainly interested in studying the effects of
first order terms.\par

We coupled the integration of the Lorentz force and the Coulomb
drag in the semi-implicit integrator of BS2010, which is very
similar to the Boris integrator \citep{Boris1970} implemented by
\citet{2009ApJ...707..404L}. Following the notation by BS2010, the
equations to solve for the particle dynamics are:

\begin{equation}
  \frac{d {\bf x}}{d t} = {\bf v}, ~~ \frac{d {\bf v}}{dt} =
       {\bf a}[{\bf v},{\bf x},{\bf u}^{n + 1/2}({\bf x})]
       \label{eq:set_dust_eqs}
\end{equation}

where we have grouped all the acceleration terms in eq. (\ref{eq:dust_eom})
under the term ${\bf a}$. Then, the velocity update can be
computed using the predicted particle position ${\bf x'}$ as:

\begin{equation}
  {\bf v}^{(n+1)} = {\bf v}^{(n)} + h \Lambda^{-1}
  {\bf a}({\bf v}^{(n)},{\bf x'},{\bf u}({\bf x'})),
  ~ \Lambda =  1 - \frac{h}{2} \frac{\partial {\bf a}}{\partial {\bf v}}
\end{equation}

where the Jacobian matrix for eq. (\ref{eq:dust_eom}) is:

\begin{equation}
  \frac{\partial {\bf a}}{\partial {\bf v}}=
  \pmatrix{
    -\frac{1}{t_{s}}-\nu _{C} &  Q_{m}B_{z} &  -Q_{m}B_{y} \cr
    -Q_{m}B_{z} & -\frac{1}{t_{s}}-\nu _{C} & Q_{m}B_{x}  \cr
     Q_{m}B_{y} & -Q_{m}B_{x} & -\frac{1}{t_{s}}-\nu _{C}
    }
  \label{eq:complete_jacobian}
\end{equation}

In the above matrix, we have renamed the term $Z_{d}e/m_{d}c$
as $Q_{m}$ for simplicity. \par

The last consideration for the implementation of
charged particles is the numerical timestep. In order
to integrate accurately the oscillation of a charged
particle, the timestep
has to be considerably smaller than the particle's Larmor
time. As a compromise between numerical accuracy
and computational cost, we have chosen the same value
as that by \citet{2017MNRAS.469.3532L}:

\begin{equation}
  \Delta t = \frac{\rm CFL}{10}\min \bigg(\frac{m_{d}c}{Z_{d}e|{\bf B}|},\Delta t_{\rm MHD}\bigg)
  \label{eq:modif_timestep}
\end{equation}

where CFL is the Courant-Friedrichs-Lewy number and $t_{\rm MHD}$ is the
timestep computed without considering the dust particles.

\section{Simulations of a Molecular Cloud envelope}\label{sec:simulations}

\subsection{Numerical model}\label{subsec:numerical_model}

We solve with Athena the equations for isothermal, ideal MHD
in 2D together with Eq. \ref{eq:dust_eom} for the dust
dynamics in a box of length $L$ with periodic boundary
conditions. \par

As a first step in the study of the dynamics of charged dust
in MCs, we developed a very simple model of a MC envelope
that responds to the common physical parameters of the WNM.
The ambient gas is composed by HI at $T = 6000 K$ with an
ionization fraction $\chi = 0.1$; the magnetic field
strength is chosen to be $B = 10^{-6}$ G, in concordance
with the measured values in the ISM of  a few $\mu$m
\citep{2001RvMP...73.1031F}. The gas density is
$n = 10$ cm$^{-3}$, a
selection justified by observational
evidence \citep{2007prpl.conf...81B,2009ApJ...705..144F,2017PASJ...69L...5F}. \par

The dust population is represented by
80,000 test particles that share the
same properties: a radius of $a_{d} = 5\times 10^{-6}$cm
and an internal density of
$\rho^{int}_{d} = 1$g cm$^{-3}$. They are randomly
distributed over the
whole domain with a velocity taken from a Gaussian
distribution with deviation $v_{th}$, the isothermal
sound speed.
For these grains, we
computed an
average electric charge
under the assumption of statistical
equilibrium of $Z = -17$ following the procedure
described in \citet{2001ApJS..134..263W}; the
  intensity of the interstellar radiation field
  \citep{1983A&A...128..212M} is considered to be $G = 1$. \par

 Initially, the medium is uniform in density, with the magnetic
 field parallel to the x-axis. The box size is
 chosen to be $L = 1$ pc, a  value that allows to
 resolve the apparent
 characteristic molecular gas filament widths of 0.1
 pc \citep{2011A&A...529L...6A} and, of more
   relevance for the present study, the width
   of HI fibers observed in the diffuse ISM
   (sizes below 0.04 pc, \citealt{2014ApJ...789...82C}):
 at the chosen
   resolution ($1024 \times 1024$, see below), one pixel size
 corresponds to $1/1024 \sim 9.76 \times 10^{-4}$ pc.
 We introduced a turbulent
 spectrum in velocity as the superposition
 of four waves with wavelengths $\lambda_{0} = L/24$,
 $\lambda_{1} = 26L/72$, $\lambda_{2} = 49L/72$, $\lambda_{3} = L$ and
 wave amplitude $u_{0} = v_{th}(k_{n}/k_{0})^{3/2}$:

 \begin{eqnarray}
   u_{x} = u_{0}\sum_{n=0}^{3}\cos(k_{n}y) \nonumber \\
   u_{y} = u_{0}\sum_{n=0}^{3}\sin(k_{n}y)\\
   u_{z} = 0 \nonumber   
 \end{eqnarray}

 The wave spectrum has been chosen such that the number
 of waves ($N_{waves} = 4 $)
 is large enough to produce a turbulent state in the gas, but
 small enough to follow the individual effects of each wave.
 The longest wavelength is chosen to be $\lambda_{max}= L$, while
 the shortest one is $\lambda_{min} = 1/24$, much larger than
 the resolution of the domain; the other wavelengths have been
 chosen from the following relationship:
 $\lambda_{n} = \lambda_{min} + n * (\lambda_{max} - \lambda_{min})/(N_{waves} - 1)$. \par
 
 The simulation has been run with a resolution of
 $1024  \times 1024$ pixels$^{2}$ and up to $t = 5$ (code units, $\sim 0.4 Myr$).
 The reason behind this choice is that at $t = 5$, the initial perturbations have been damped by a factor of two and they are bounded for $t>5$,
 so we consider that the system has been stabilized.

\subsection{Decoupled motions of gas and dust}\label{subsec:gas_dust}
Motivated by the apparent linear relationship between
gas and dust column density reported by
several authors
\citep{1977ApJ...216..291S,1978ApJ...224..132B,1995A&A...293..889P,2017ApJ...846...38L}, we studied the
distribution of gas and dust in the simulation. \par

\begin{figure}[h!]
  \plotone{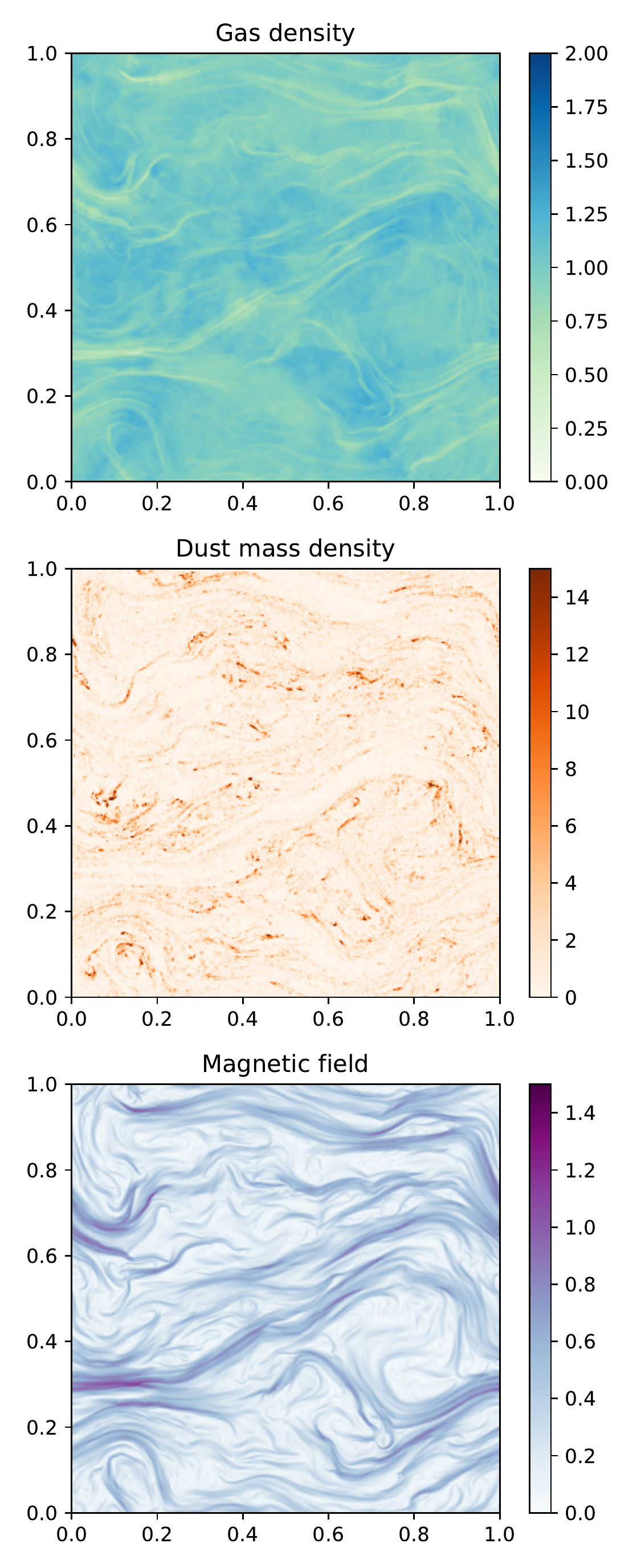}
  \caption{\textit{Upper panel}: gas density map
    at the end of the simulation. It is normalized such that at t = 0,
    the density is constant and equal to 1. \textit{Middle panel}:
    dust mass density map (see the text for the details of how it was
    built) for the same stage of the simulation. A dust mass
    value of 1 corresponds to the assumed dust-to-gas ratio of 0.01. Some
    clumps with larger values are observed all over the domain.
    \textit{Lower panel}: magnetic field modulus, normalized such
      that at t = 0, it has a uniform value of $\sim 0.22$. Note the
    correlations between the dust and magnetic field structures.
  }
  \label{fig:dust_gas_map}
\end{figure}

We built dust distribution maps from the
particle's positions assuming an initial
dust-to-gas ratio $\rho_{d}/\rho_{g} = 0.01$
\citep{1954ApJ...120....1S}. For the initial
conditions, we estimated the dust mass contained in the total
volume of 1 pc$^{2} (\times 1$ pc) and distributed it equally
over the 80,000 test particles. We considered
that the test particles
trace the distribution of real particles, too large to
  be computationally resolved, and that those real
  particles are uniformly distributed around the test ones. This
  post-processing is done after the main simulation, so it does not
affect the formation of dust filaments.
We distributed the dust mass over a matrix with the
same size as the resolution of the simulation ($1024 \times 1024$) and
normalized it such that a value of 1 corresponds
to the assumed dust-to-gas ratio of 0.01.
The dust and gas
maps at the final stage of the simulation
are shown in Fig. \ref{fig:dust_gas_map}.

\begin{figure}[h!]
  \plotone{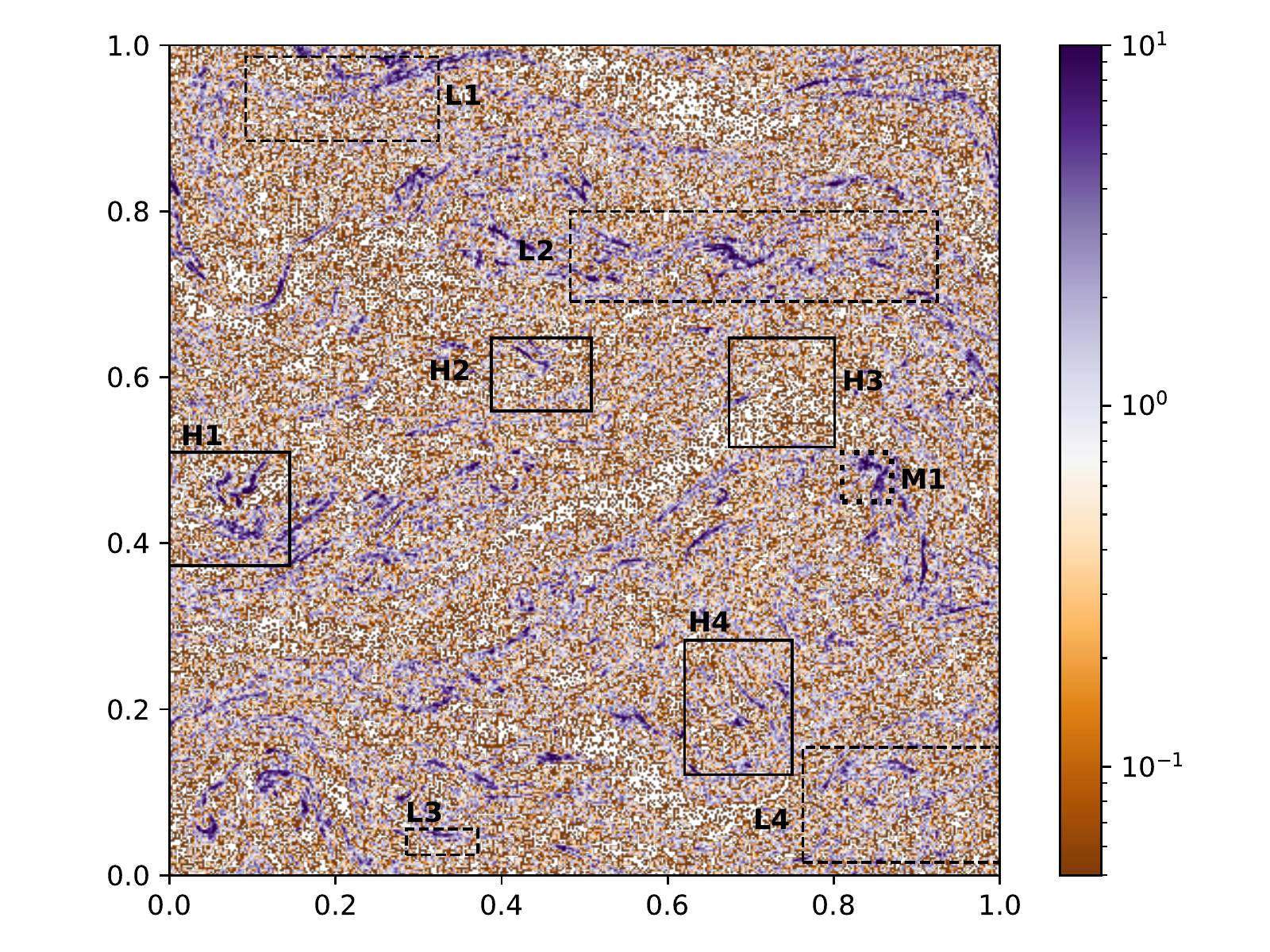}
  \caption{Dust-to-gas ratio
    map for the final stage of the simulation \textbf{in log-scale}. A value
    of 1 \textbf{(0 in log scale, white colors)}
    corresponds to the standard ratio of 0.01; larger values
    reveal regions where the dust density is dominant, while the
    regions with a value of 0 correspond to regions with a low or
    inexistent abundance of dust. Thick (dashed) rectangles encloses regions
    of high (low) gas density, and the small dotted region encloses
    an overdensity of dust that corresponds to a region of mid density.
  }
  \label{fig:dgratio_5}
\end{figure}

If we look at the dust-to-gas ratio maps in Fig. \ref{fig:dgratio_5}, we
observe that there are certain regions where the values are significantly
larger than the initial average of 0.01 (value equal to unity
in the normalized maps). On the other hand, there are some regions
of the simulation where the dust-to-gas ratio is very small or
that are completely devoid of dust. This is an expected result of
  the submission of charged dust to a spectrum of Alfv\'en waves. Finally, it is worth noting
that the dust
overdensities show a filamentary structure that
will be studied in Sec. \ref{subsec:dust_filaments}.

\begin{figure}[h!]
  \plotone{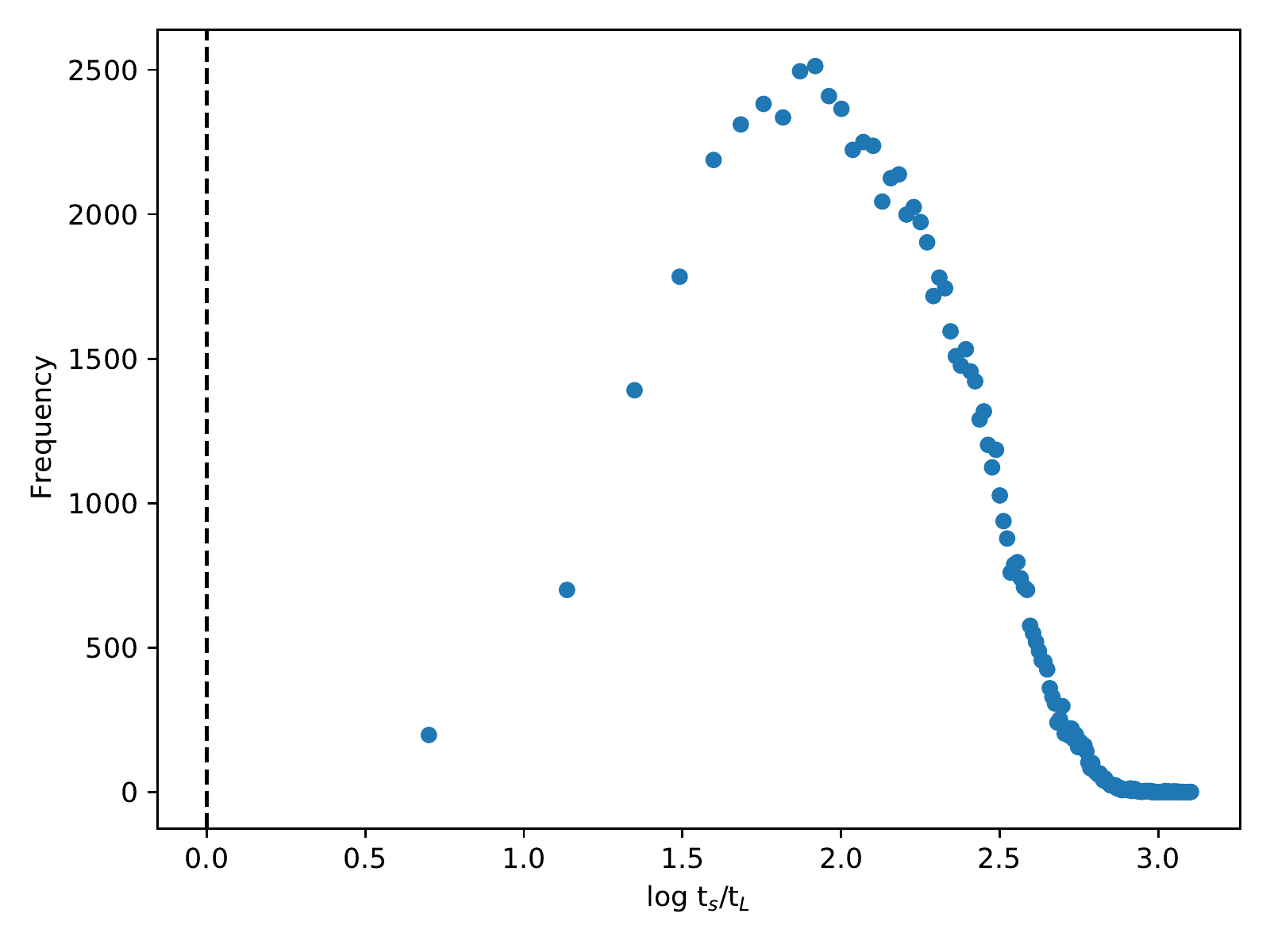}
  \caption{Histogram of the ratio $t_{s}/t_{L}$, particle
    stopping time over Larmor time (in logarithmic scale), for the final stage
    of the simulation. All particles show values
    greater than 0 ($t_{s}/t_{L} = 1$, black dashed line), confirming
    the magnetic field coupling.\label{fig:ts_tL}}
  
\end{figure}

The decoupling of gas and dust could arise
due to the large differences between the particle stopping
time $t_{s}$ and the Larmor time, $t_{L} = m_{d}c/|Z_{d}|e|\mathbf{B}|$.
For all particles in the simulation at the final stage, the ratio
$t_{s}/t_{L}$ is always greater than one (see Fig. \ref{fig:ts_tL})
indicating that the coupling with the magnetic field is dominant. We
selected some regions in the gas density map in order to
study in detail the relationship between the distribution of
dust, the gas density and the magnetic field strength. The
regions are displayed in Fig. \ref{fig:dgratio_5} and were
selected such that a region of high (low) gas density shows values
greater (lower) than $1\sigma$, since the gas distribution
can be approximated by a Gaussian.\par

\begin{figure*}
  \gridline{\fig{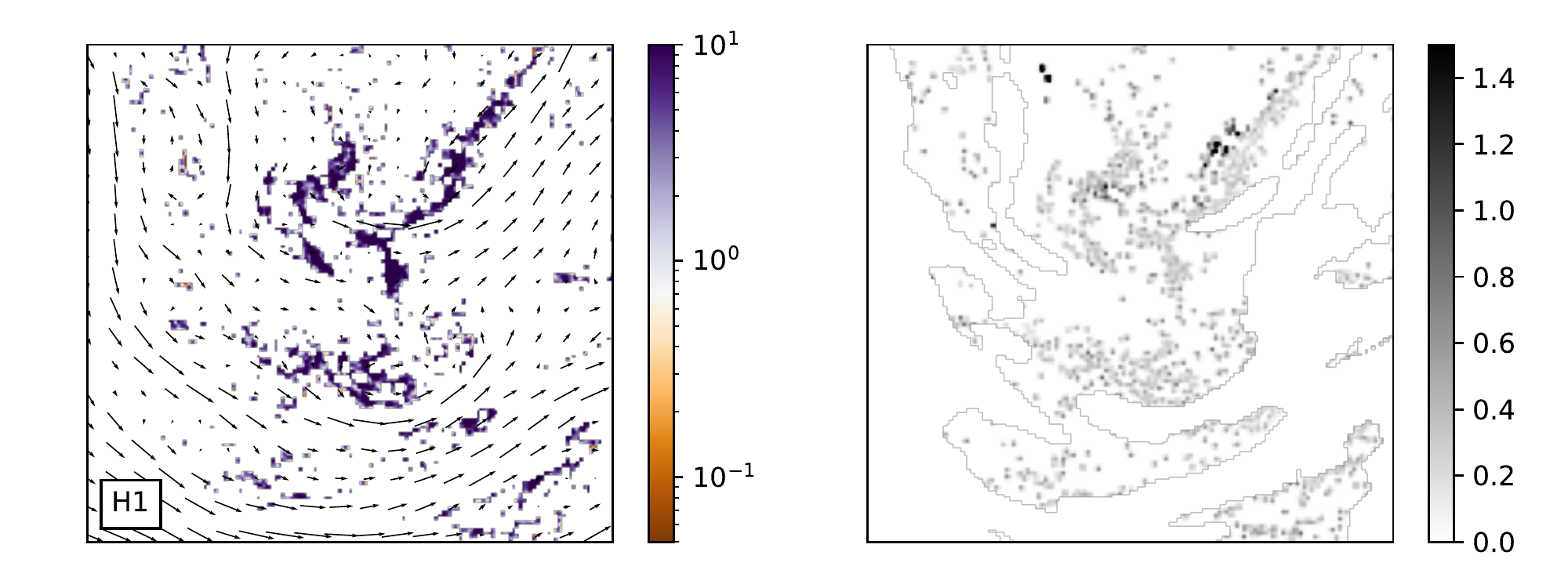}{.7\textwidth}{}}
  \gridline{\fig{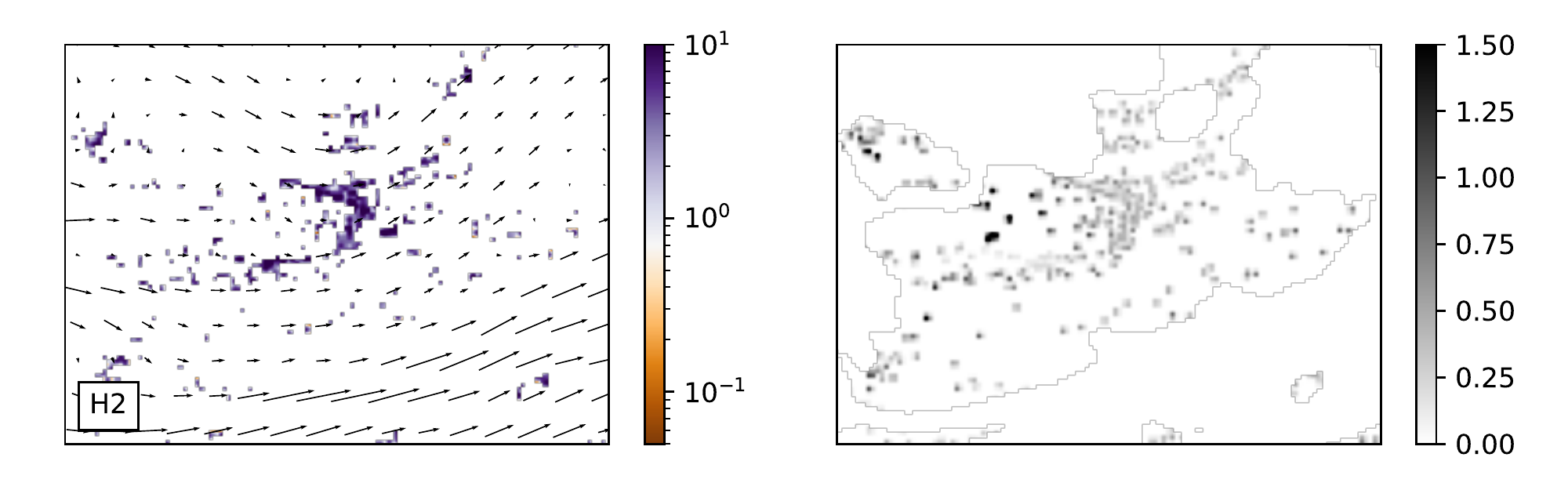}{.7\textwidth}{}}
  \gridline{\fig{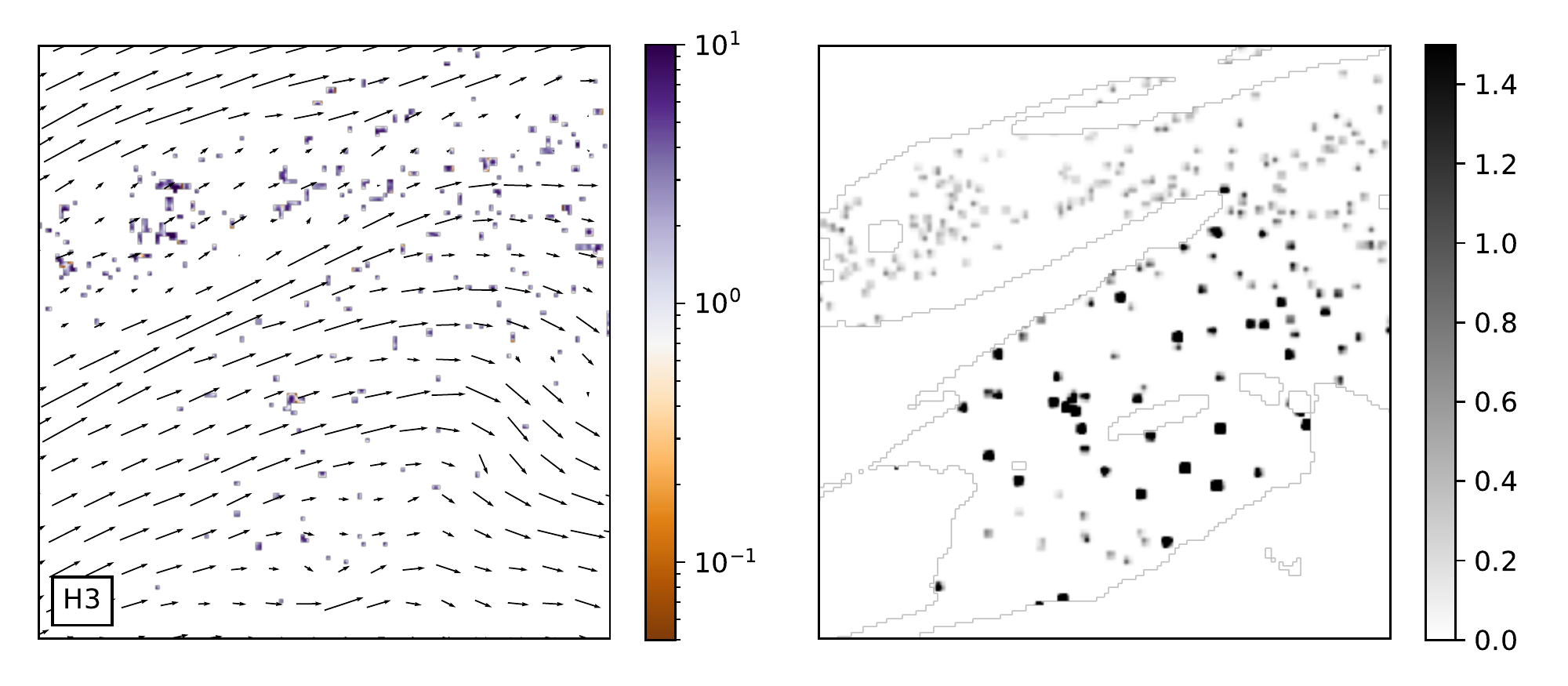}{.7\textwidth}{}}
  \gridline{\fig{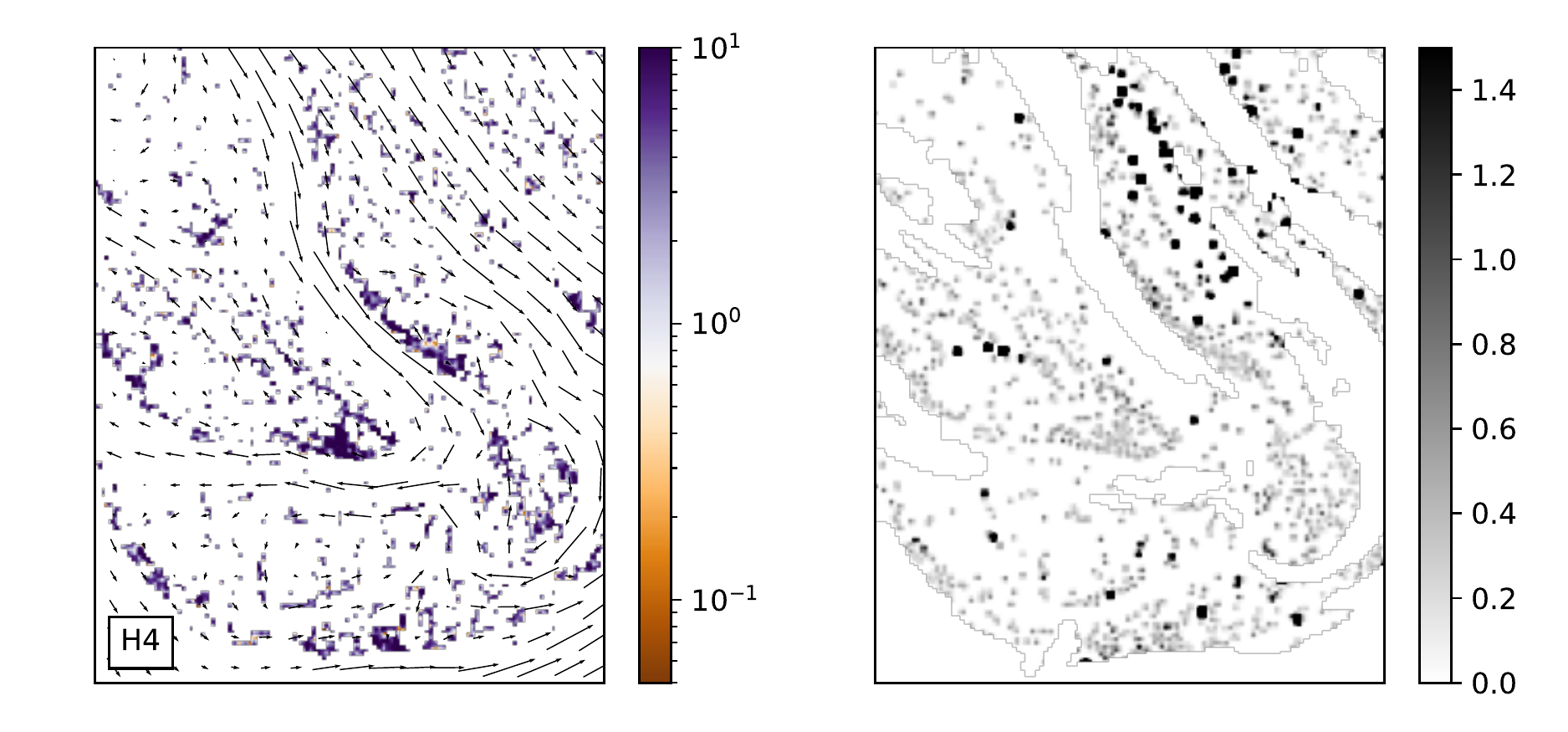}{.7\textwidth}{}}
  \caption{Zoom of four high gas density regions labeled as
    H1-H4. On each panel, the left coloured map represents
    the normalized dust-to-gas ratio as in Fig. \ref{fig:dgratio_5}
    with the magnetic field overlaid. Right panels
    represents the dust-to-gas velocity ratio.\label{fig:H1-H4}}
\end{figure*}

\begin{figure*}
  \gridline{\fig{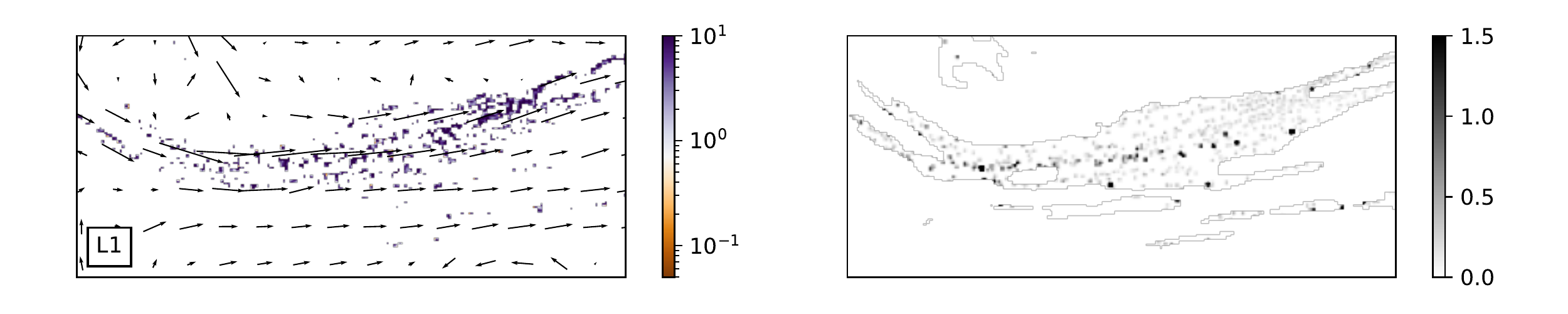}{\textwidth}{}}
  \gridline{\fig{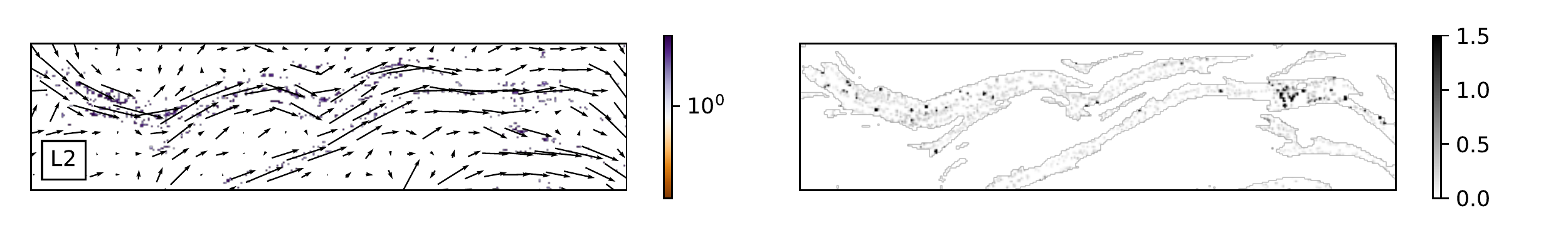}{\textwidth}{}}
  \gridline{\fig{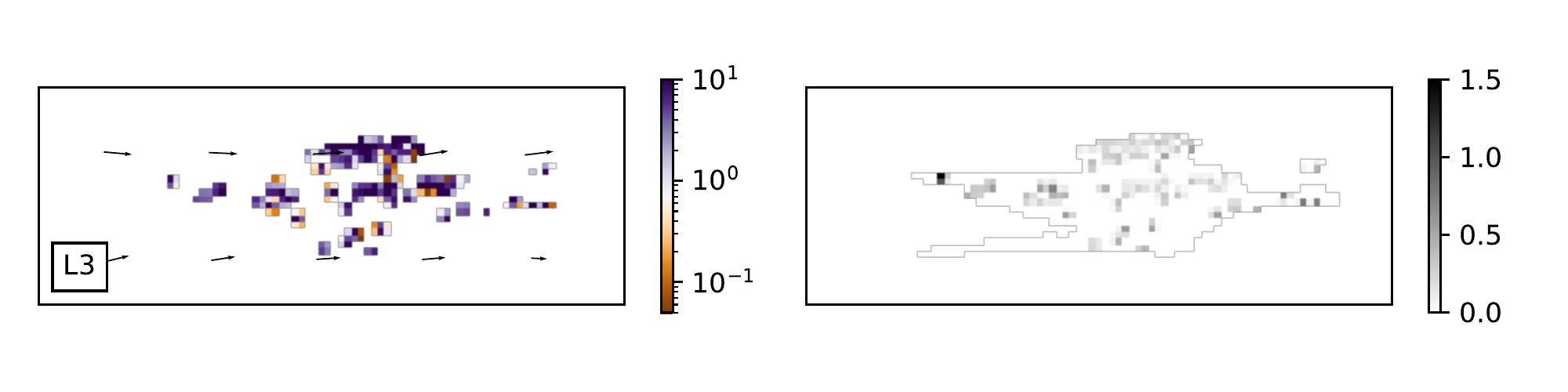}{\textwidth}{}}
  \gridline{\fig{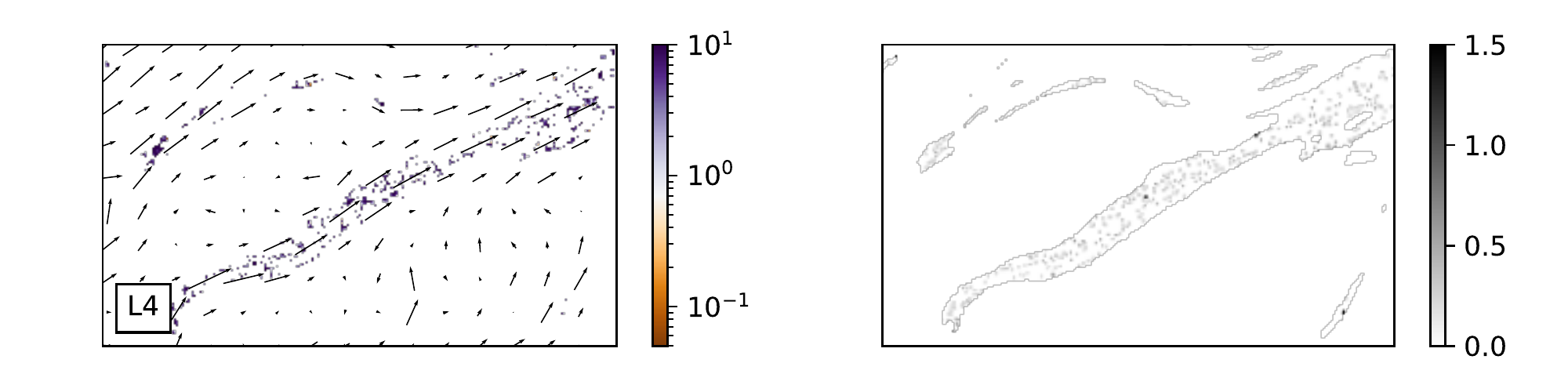}{\textwidth}{}}
  \caption{Same as Fig. \ref{fig:H1-H4} but for low
    gas density regions.\label{fig:L1-L4}}
\end{figure*}

\begin{figure*}
  \gridline{\fig{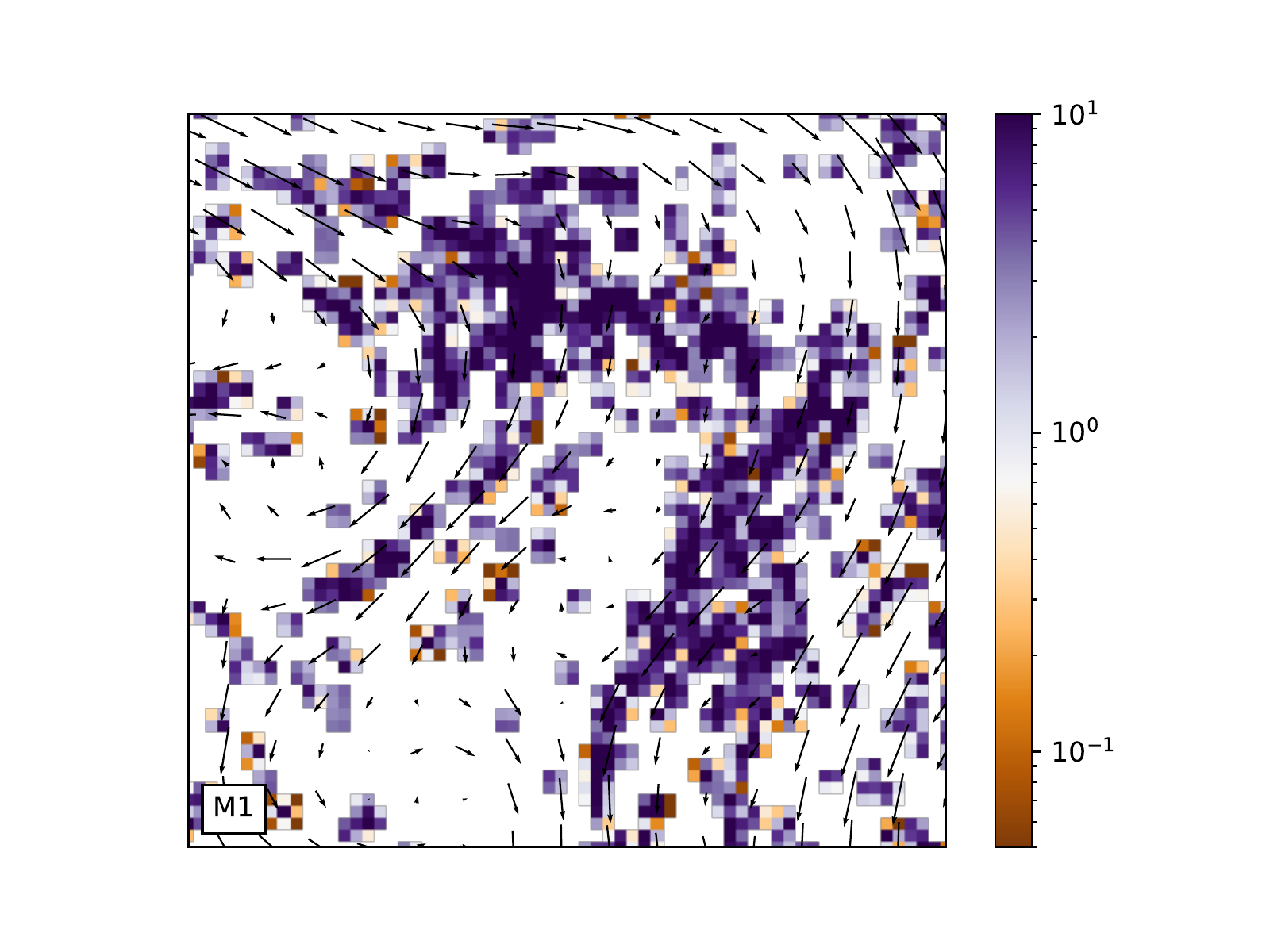}{.5\textwidth}{}
  \fig{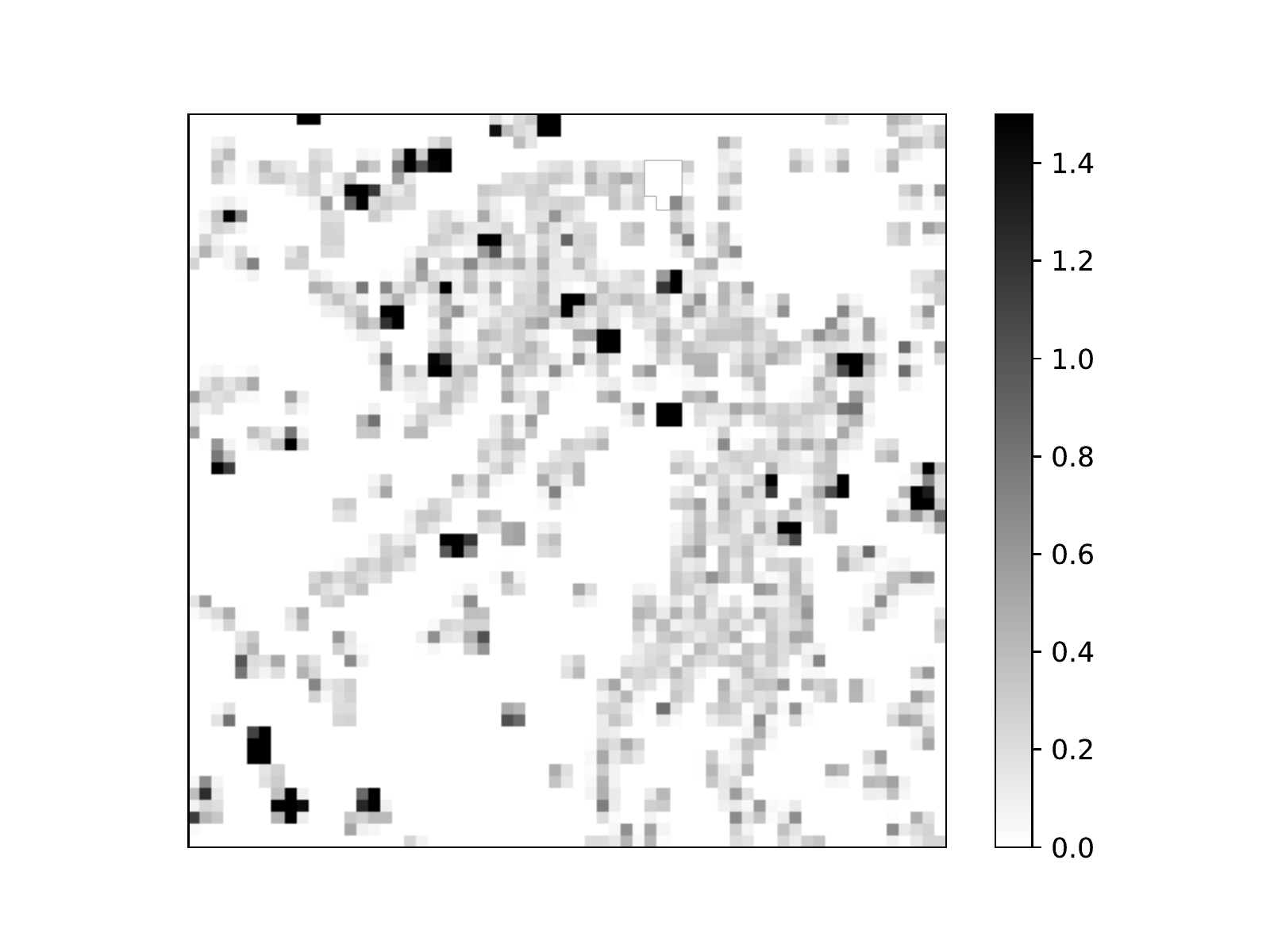}{.5\textwidth}{}}
  \caption{Same as Fig. \ref{fig:H1-H4}, but for a gas mid-density
    region. \label{fig:M1}
  }
\end{figure*}

The analysis of the selected regions shown in  Fig. \ref{fig:dgratio_5}
may be summarized as follows:

\begin{itemize}
  
\item At high gas density regions (Fig. \ref{fig:H1-H4}) we
observe large values of the dust-to-gas ratio forming
filaments, parallel to the magnetic field.

\item Except
for region H3, where the dust density is extremely poor,
the dust mean velocity is similar to the Alfvén velocity,
pointing toward a strong coupling of dust with
the magnetic field.

\item At low gas density regions
(Fig. \ref{fig:L1-L4}), the dust filamentary structure
is still observed, but there is more granulation and the
filament widths are narrower. The modulus of the 
magnetic field in these
regions is approximately 2.5 times greater than in
regions H1-H4, and the dust velocity is again
similar (and for regions L1 and L4 almost equal) to the
Alfvén velocity.

\item Finally, for the mid gas density
region (Fig. \ref{fig:M1}) we observe a hybrid behavior.
Dust filaments are also formed parallel to the magnetic
field, but the overdensities are broader and,
generally, perpendicular to the field. In this case,
the mean dust motion seems more coupled
to the gas dynamics.

\end{itemize}

\subsection{Filamentary dust structure}\label{subsec:dust_filaments}

\begin{figure}[h!]
  \plotone{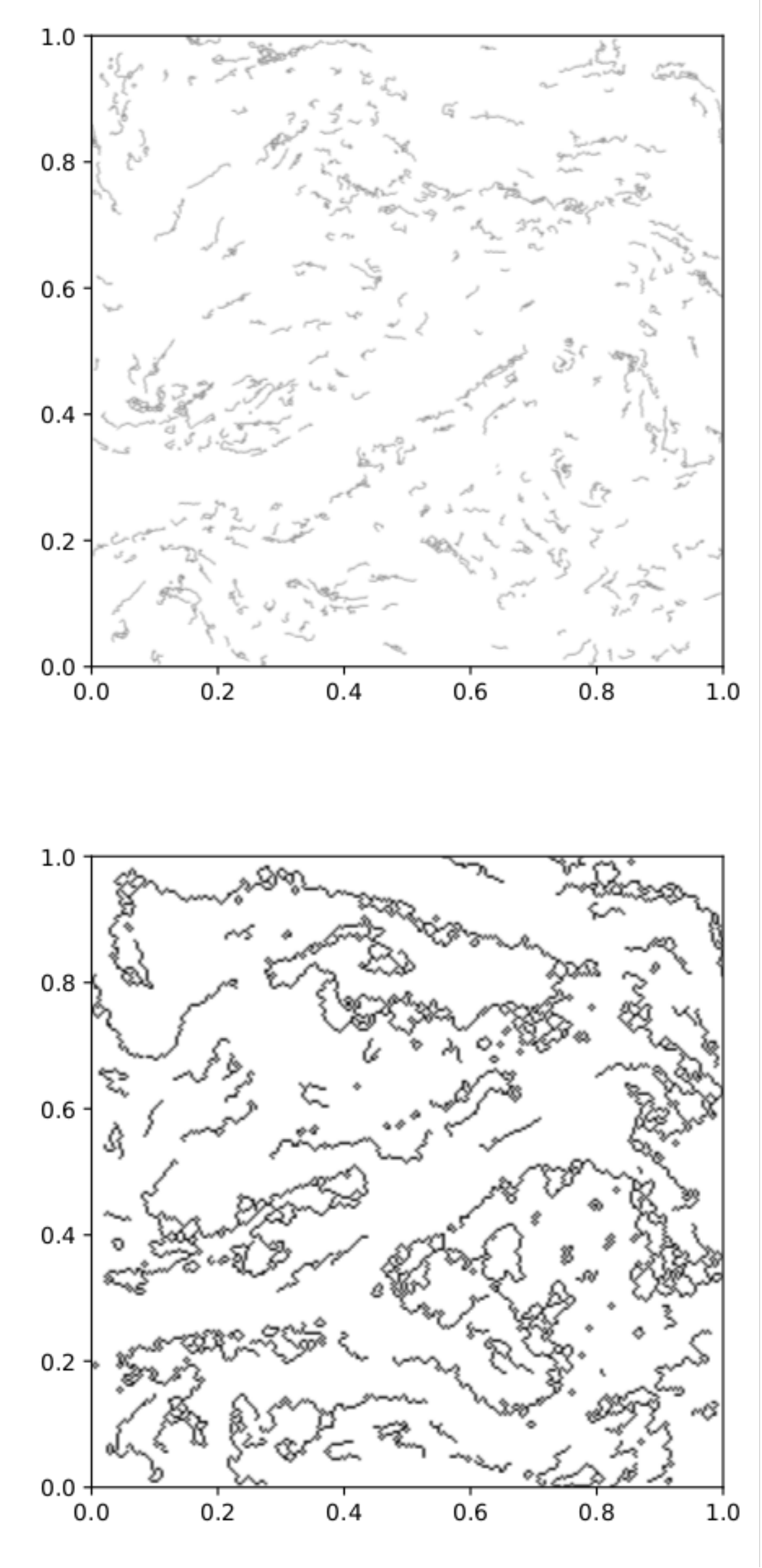}
  \caption{
    Dust filaments found by FilFinder for the
    original dust mass map (upper panel) and
    for a degraded version with one fourth
    of the resolution (lower panel). Short and numerous
    filaments are found for the original image, corresponding
    to the dense clumps. For the degraded image, longer
    branches tracing the general shape of the filaments are
    observed.
        \label{fig:skeletons}
  }
  
\end{figure}

In order to better understand the properties
of the dust filaments generally decoupled from
the gas (see Sec. \ref{subsec:gas_dust}), we analyzed
in detail their properties using FilFinder
\citep{2015MNRAS.452.3435K} over the
dust mass map at the final stage of the
simulation. For the
analysis, we created a custom mask
preprocessing the map with the
python
packages scikit-image \citep{scikit-image}
and scikit-learn \citep{scikit-learn}. First, we binarized the
map considering only pixels with a dust mass greater
than the median; then, we identified connected regions
and removed those with less than 10 pixels. The
resulting mask is then passed as an argument to
FilFinder together with the original
dust map; we then skipped the flattening and creation
of a mask steps and performed directly the identification
and pruning of filaments; the results are
shown in Fig. \ref{fig:skeletons}.

We found useful to study the filamentary structure over
a degraded version of the map ($256\times 256$ px$^{2}$) as well
as over the original
one ($1024 \times 1024$ px$^{2}$).
While the latter represents well the fragmented
structure of the dust clumps, the former gives an overall
picture of the morphology of the dust filaments. The typical
length of the filaments in the original image
is 0.04 pc, while for the degraded image is 0.21 pc.
We also computed the typical widths of the filaments
as the mean width perpendicular to the longest
path and found a value of 0.0035 pc for the
high resolution image and 0.0087 pc for the degraded one;
note that the image size imposes a resolution
  limit of $9.8 \times 10^{-4}$ pc for the $1024 \times 1024$ px$^{2}$
  map, and of $3.9 \times 10^{-3}$ pc
  for the $256 \times 256$ px$^{2}$ one, but
  both mean widths are above that threshold.
Table
\ref{tab:params} summarizes all the parameters
of the analyzed regions.
It
is evident that, although the resolution of the
dust map critically affects the filament length, the widths
are similar and  an order of magnitude smaller than
the typical width of 0.01 pc for gas filaments in molecular clouds
found by Herschel (\citealt{2011A&A...529L...6A} but
 see \citealt{2017MNRAS.466.2529P} for a recent discussion on the validity of
 such a characteristic width); since at large scales
 the correlation between dust and gas still holds,
 our results are in concordance with
 the observational report by \citet{2014ApJ...789...82C} on diffuse
 HI filaments at all resolved scales, up to a lower limit of 0.04 pc imposed by
 the instrumental resolution.
 However, the dust filament width should not depend
  on the resolution of the global simulation, because dust
  dynamics is resolved based on the gas properties on
  the adjacent cells. Hence, as dust grains are coupled to the magnetic
  field, an increase of the resolution only contributes to a
  better definition of the shape
  of the gas filaments, but the overall dynamics of dust grains should
  remain unchanged. What does affect the morphology of the dust filaments
  is the wave spectrum, as is discussed later in the following section.
  The dust structures
  will be more clumpy or filamentary depending on the propagating waves.
  In that sense, the box size may
  play a role since  it sets an upper limit to the longest
  wavelength. \par

\begin{deluxetable*}{ccccccccc}
\tablecaption{Properties of the analyzed regions\label{tab:params}}
\tablewidth{0pt}
\tablehead{
\colhead{Region} & \colhead{Dimensions} & \colhead{$<DGR>$} & \colhead{$<B>$} & \colhead{$<v_{A}>$} & \colhead{$<v_{gas}>$} & $<v_{dust}>$ & w$_{1024}$ & w$_{256}$\\
\colhead{} & \colhead{pc $\times$ pc}   & \colhead{} & \colhead{$10^{-6}$ G}
& \colhead{$10^{5}$ cm s$^{-1}$} & \colhead{$10^{5}$ cm s$^{-1}$} & \colhead{$10^{5}$ cm s$^{-1}$} & \colhead{$10^{-3}$ pc} & \colhead{$10^{-3}$ pc}}
\startdata
H1 & 0.145 $\times$ 0.138 & 1.403 & 0.737
& 0.476  & 2.997 & 0.205 & 4.385 & 10.449\\
H2 & 0.121 $\times$ 0.089 & 0.798 & 0.829
& 0.534 & 2.402  & 0.135 & 2.981 & 9.975\\
H3 & 0.128 $\times$ 0.133 & 0.280 & 1.262
& 0.806 & 3.069 & 0.361  & 3.818 & 6.331\\
H4 & 0.131 $\times$ 0.163 & 1.090 & 1.015
& 0.649 & 2.629 & 0.558  & 4.109 & 9.827\\
L1 & 0.233 $\times$ 0.103 & 1.597 & 2.703
& 2.124 & 6.293 & 0.479 & 3.046 & 8.523 \\
L2 & 0.443 $\times$ 0.109 & 1.767 & 2.437
& 1.841 & 5.935 & 0.597 & 3.300 & 11.728 \\
L3 & 0.087 $\times$ 0.032 & 1.921 & 1.862
& 1.368 & 1.982 & 0.162 & 4.112 & 8.464 \\
L4 & 0.238 $\times$ 0.140 & 1.459 & 2.514
& 1.910 & 5.010 & 0.280 & 3.362 & 10.919\\
M1 & 0.062 $\times$ 0.060 & 2.514 & 0.777
& 0.532 & 6.104 & 1.534 & 4.939 & 17.578
\enddata
\tablecomments{Mean quantities for the dust-to-gas ratio
  (denoted by $DGR$), magnetic field, Alfvén velocity,
  gas velocity, and dust velocity. Also included are the mean
  dust filament widths on each region for the original
map (w$_{1024}$) and for a degraded version (w$_{256}$).}
\end{deluxetable*}

\begin{figure}[h!]
  \plotone{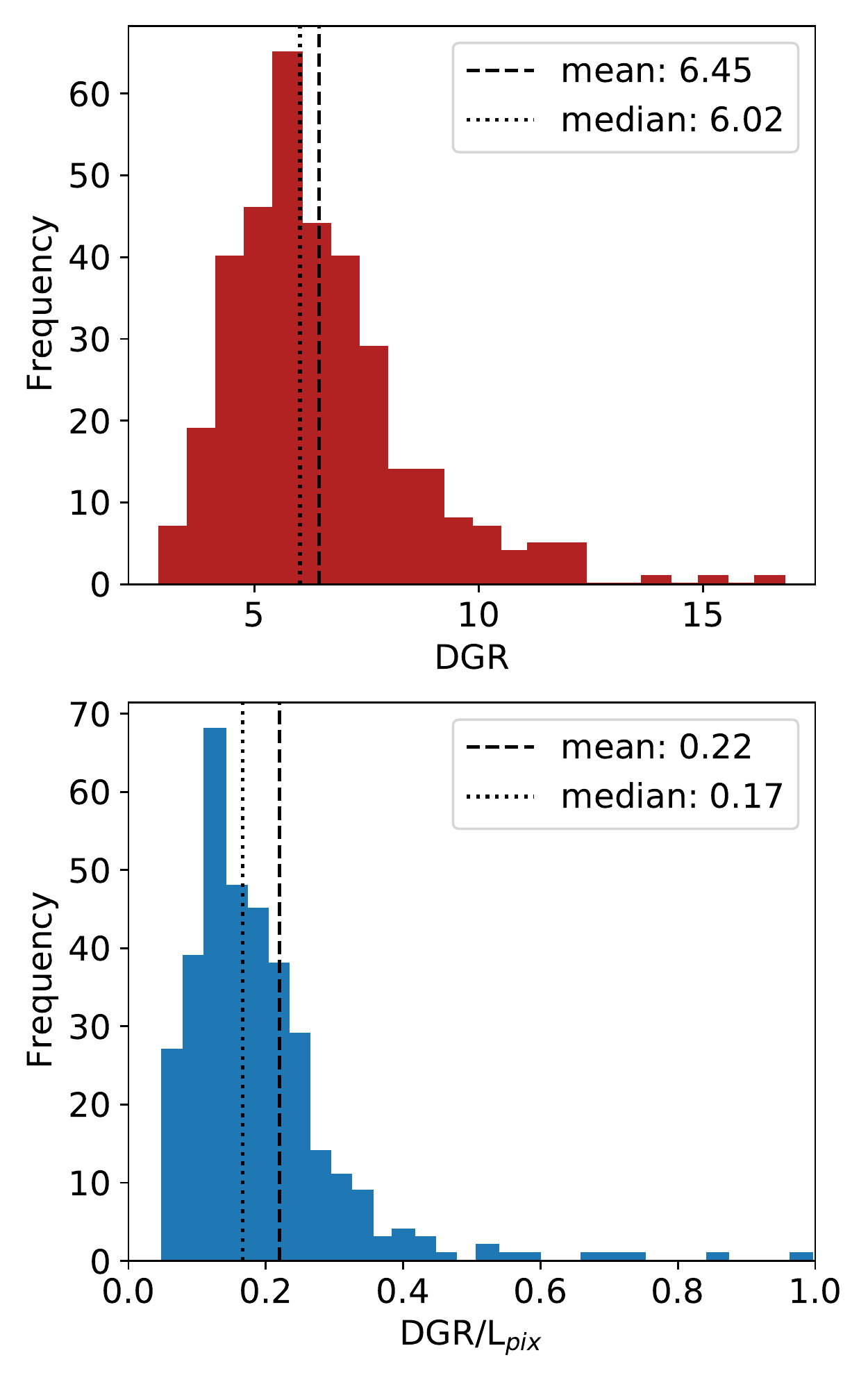}
  \caption{ \textit{Upper panel}: histogram of the
    dust-to-gas ratio along the filaments found by
    FilFinder over the original dust map. The values
    range between 3 and 17 times the typical
    ISM ratio (value of 1, see Fig. \ref{fig:dgratio_5}).
    \textit{Lower panel}: dust-to-gas ratios normalized
    by their length in pixels, L$_{pix}$. 
  \label{fig:fils_dgr}}
\end{figure}

 We have also studied the mean dust-to-gas ratio along
  the longest paths of the filaments, and the results are
  shown in Fig. \ref{fig:fils_dgr}. We find that, in
  general, the ratio reaches values six times greater than
  the average value for the ISM and spans from $\sim 3$
  to $\sim 17$, another indicative that
  dust filaments are not tightly correlated with gas filaments.
  However, if those mean values are normalized by the
  filament's length (number of pixels, L$_{pix}$,
  Fig. \ref{fig:fils_dgr}, lower panel) one can see that
  the dust-to-gas ratio inside each filament is approximately
  constant and proportional to its length by a factor that oscilates
  between 0.1 and 0.3. There not seems to be any apparent
  relationship between the dust-to-gas ratio of a dust filament
  and its width, what suggests that such width may be characteristic
  of the dust filaments. However, to confirm that hypothesis, more
  realistic simulations that consider a whole representative dust
population are needed.

\begin{figure}[h!]
  \plotone{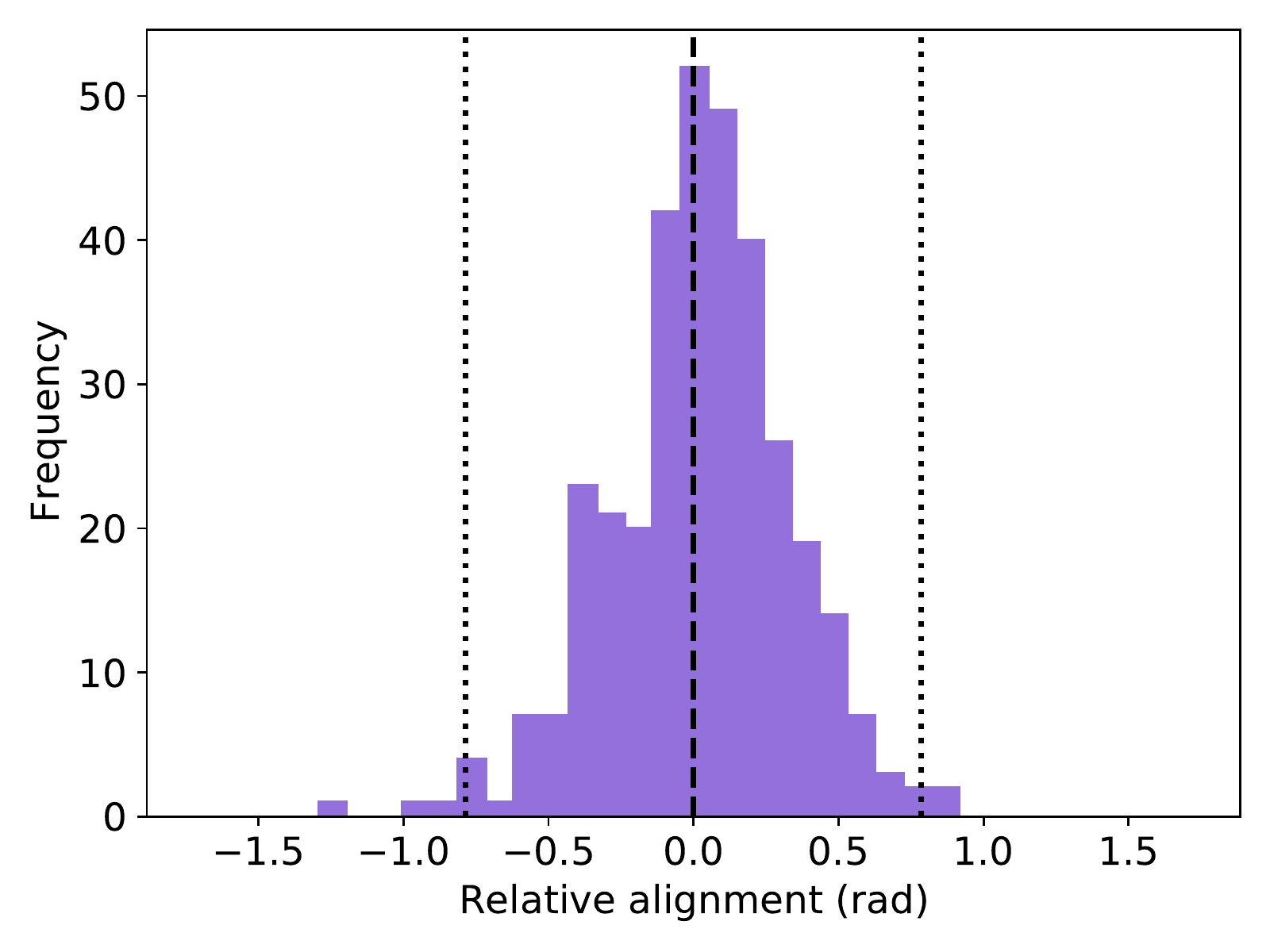}
  \caption{Histogram of relative alignment between the
    dust filaments and the magnetic field. The dashed vertical
    line is plotted at 0 rad (perfect parallel alignment)
    while the vertical dotted lines correspond to
    $\pm \pi/4$ rad.
  \label{fig:fil_alignment}}
\end{figure}

 With respect to the relative orientation between the
  dust filaments and the magnetic field, we find that
  in general dust filaments tend to be aligned with the magnetic
  field, see Fig. \ref{fig:fil_alignment}; in the
  high-resolution map, $\sim 15$\% of the filaments have relative orientations
  between $\pm 0.05$ rad (parallel), while the vast
  majority (98\%) have relative angles inside the range $\pm \pi/4$
  (dotted lines in Fig. \ref{fig:fil_alignment}).
  These angles have been computed along the skeleton
  of each dust filament as follows: every three pixels of the
  skeleton, the dot product between the magnetic field vector
  at the intermediate pixel and the skeleton vector is used
  to retrieve the relative angle between them, and is
  \textbf{restricted} to be inside the range $([-\pi/2,\pi/2))$.   

\section{Discussion}\label{sec:discussion}

It is a common practice to assume a standard
relationship between the dust and gas abundance. However,
local variations of the dust-to-gas ratio are
to be expected \citep{2006ApJ...649..807P,2014ApJ...797...59H}, especially
when the dust particles are charged
\citep{2002ApJ...566L.105L,2017MNRAS.469.3532L}, and they
are actually found in the diffuse medium and even inside
a cloud \citep{2017ques.workE..27S,2015MNRAS.448.2187C,2015ApJ...811..118R}.
There
were some early attempts to determine the scale
of the fluctuations from extinction measurements
\citep{1997A&A...319..948T,1999A&A...351.1051T}
and variations of the dust size distribution have
been found in MCs
\citep{2009ApJ...701.1450F, 2015A&A...579A..15K, 2015MNRAS.449.3867G,2017MNRAS.469.2531B}. However, most
of the studies are focused on the cool interiors of MCs since
the main interest is to determine the possible influence
of this phenomenon over the star formation processes.
But if one aims to understand
how dust may affect star formation and the formation
of molecular clouds themselves, it is necessary to
begin  analyzing its behaviour in MC envelopes, where
  the coupling to the magnetic field favors the propagation of
  hydromagnetic waves into the cloud and increase the turbulent state of the
molecular gas. \par

\begin{figure}[h!]
  \plotone{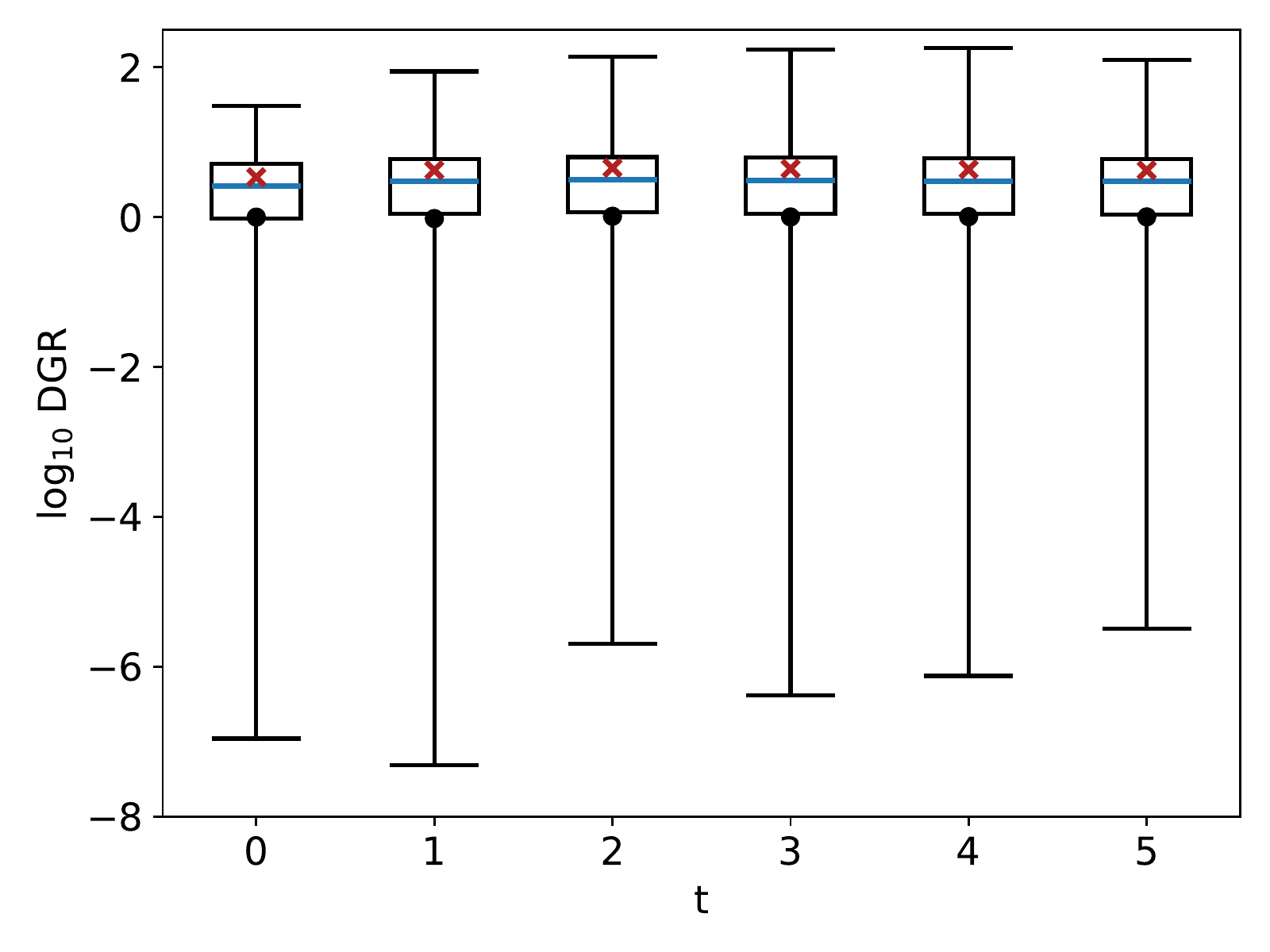}
  \plotone{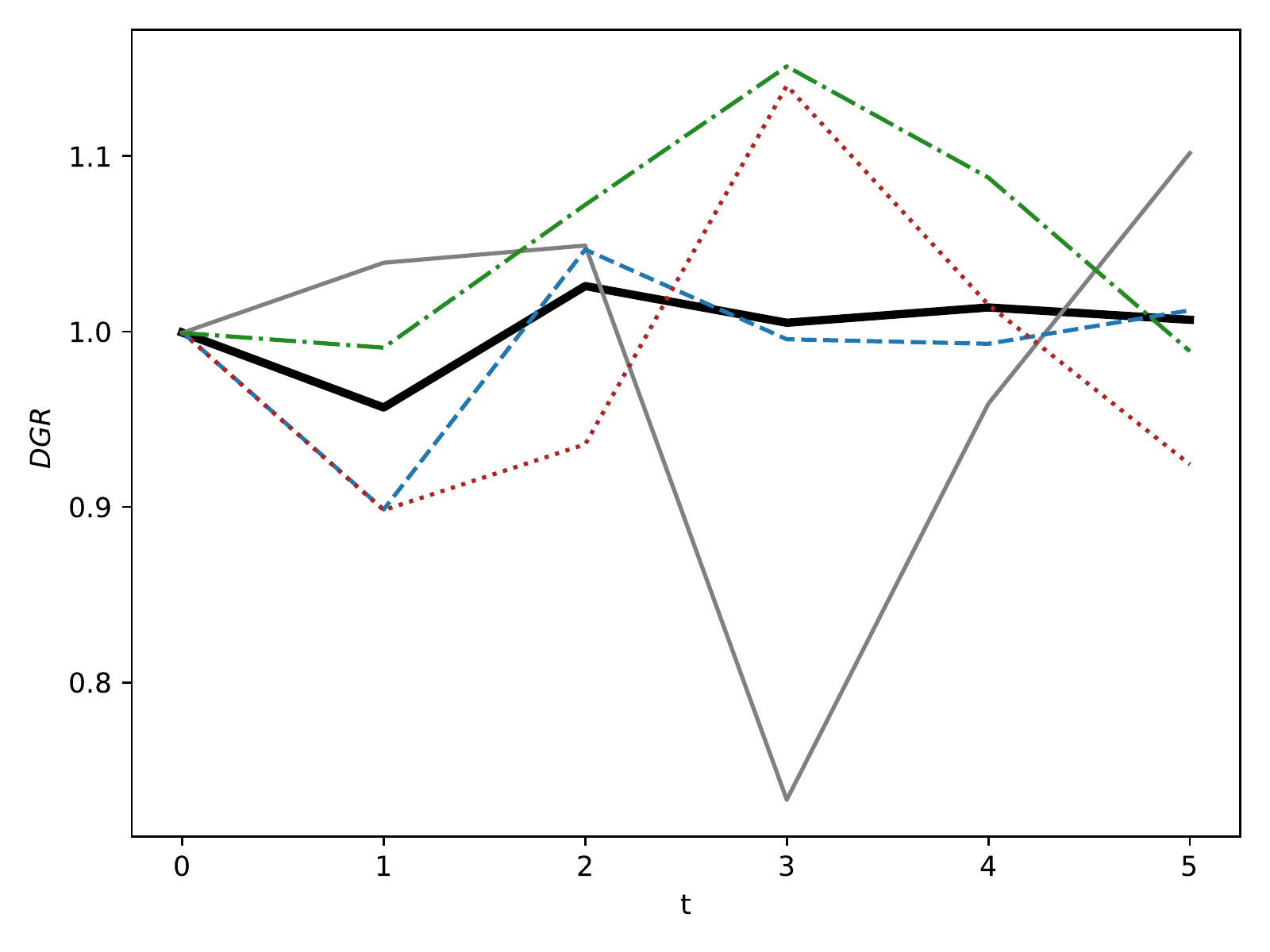}
  \caption{
    Time evolution of the dust-to-gas ratio, denoted
    by $DGR$. The upper panel shows, in logarithmic
    scale,
    a boxplot for dust-to-gas ratio values
      different from zero. The horizontal blue line
      limits the median, the mean is represented by
      a red cross, and the whiskers limits have been
      set to the maximum and minimum values. If null values
      in the map are considered, both the median and the first
      quantile are 0 (in linear scale); however, the mean
      represents well the assumed initial dust-to-gas
      ratio (black circles).
    The lower panel shows the variations
    of the mean ratio (black thick line) and the
    local variations when the whole simulation is
    analyzed in four large quadrants
    (gray solid line, blue dashed line,
    red dotted line, and green dotted dashed line).
    Note that the overall mean ratio is conserved although
    local variations are observed.
    \label{fig:dgratio_variations}
  }
\end{figure}

\begin{figure}[h!]
  \plotone{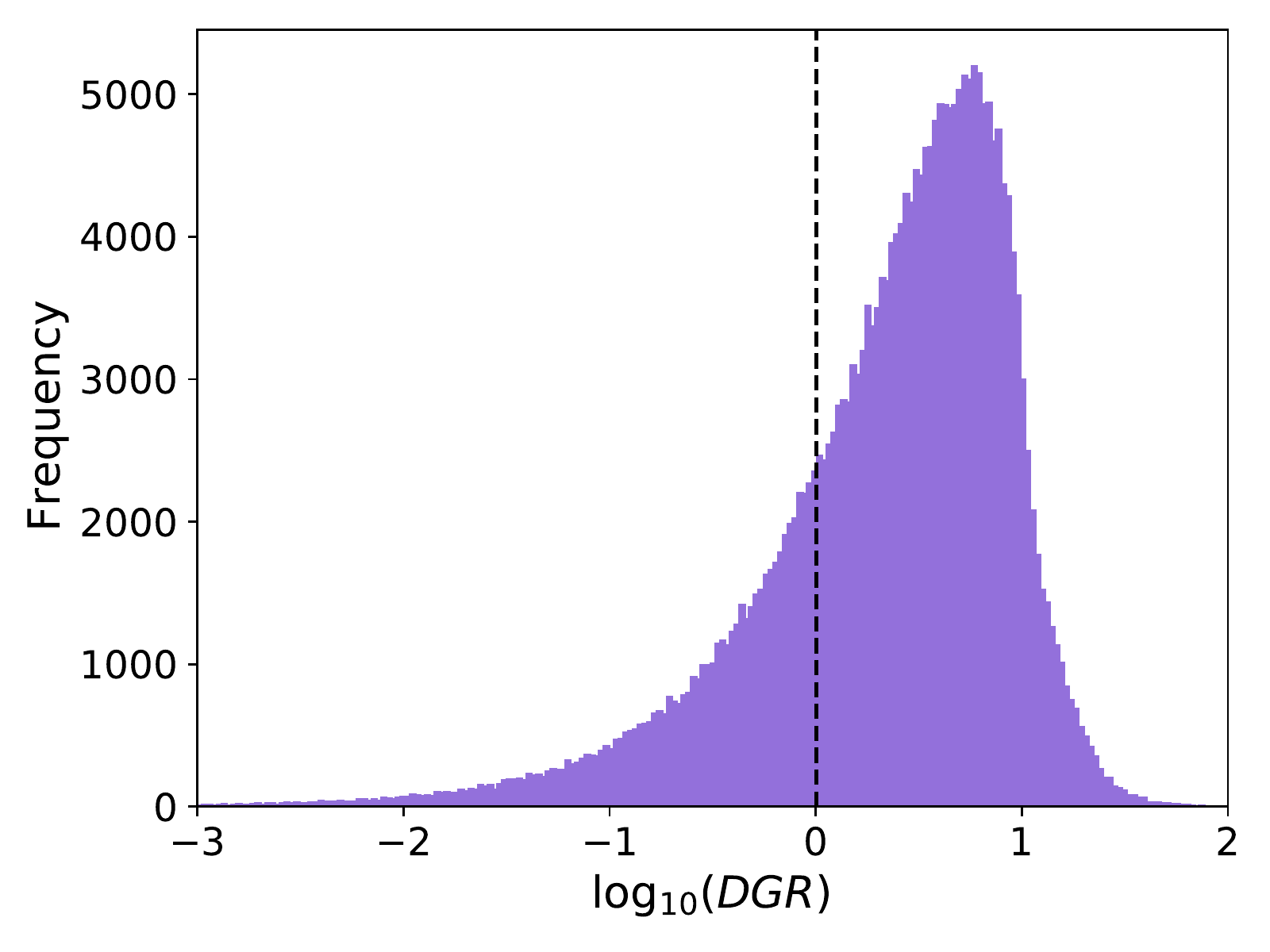}
  \caption{Histogram of the dust-to-gas ratio map shown in Fig.
    \ref{fig:dgratio_5}. Values greater than 1 (0 in the logscale,
    vertical dashed line) indicate a dust density value above
    the expected one. This plot highlights the fact that
    values ten times greater than the expected for
    the standard ratio of the diffuse ISM are common in the
    simulation, and that the relationship between
    gas and dust densities is far from being linear.
        \label{fig:dgratio_histogram}
  }
\end{figure}

Despite the simplicity of our model and having considered
an initial homogeneous distribution of dust and gas, we have shown that
the propagation of velocity waves produce local variations of the
dust-to-gas ratio, forming narrow dust filaments
aligned with the magnetic field. Due to the random distribution
of dust grains at the beginning of the simulation, the ratio
is not constant even at the initial stage of the simulation, although
the mean scarcely varies with time at large scales, as shown
in Fig. \ref{fig:dgratio_variations}. This fact could
support the argument that, at large scales, the dust-to-gas
ratio is nearly constant. However, at small scales there are
  appreciable deviations of the ratio that are far from a linear relationship.
  In Fig. \ref{fig:dgratio_histogram}, the histogram of the dust-to-gas
  ratio map presented in Fig. \ref{fig:dgratio_5} is shown; there are clear variations
  of the ratio, and the peak of the distribution is far from the assumed
dust-to-gas ratio (dashed vertical line).

  Recently, \citet{2017MNRAS.469.3532L} reported
  deviations of the dust-to-gas ratio in the cold
  neutral medium and \citet{2018MNRAS.479.4681H}
  studied in detail the spectrum of instabilities
  that may arise when charged dust moves in a
  magnetized gas. In particular, they report
  that for the WNM the relative velocity between
  dust particles and gas should be subsonic and that
  the dominant resonances tend to form structures
  parallel to the magnetic field. Our results are
  compatible up to a point, since the majority
  of the structures are indeed aligned with the
  magnetic field and the mean dust and gas
  velocities differ a
  few tens of km s$^{-1}$, see Table \ref{tab:params}.
  These results are also in good agreement with
    observational evidence of dust filaments with
    low column densities aligned with the magnetic field
    \citep{2016A&A...586A.135P,2019ApJ...878..110F}.
  There seems to be only an exception with the
    mid-density region (see Fig. \ref{fig:M1}), where
    a dust structure is formed perpendicular to the
    magnetic field direction. We attribute this localized
    phenomenon
    to the clear decrease of the
    magnetic field strength along the structure.\par

\begin{figure}[h!]
  \plotone{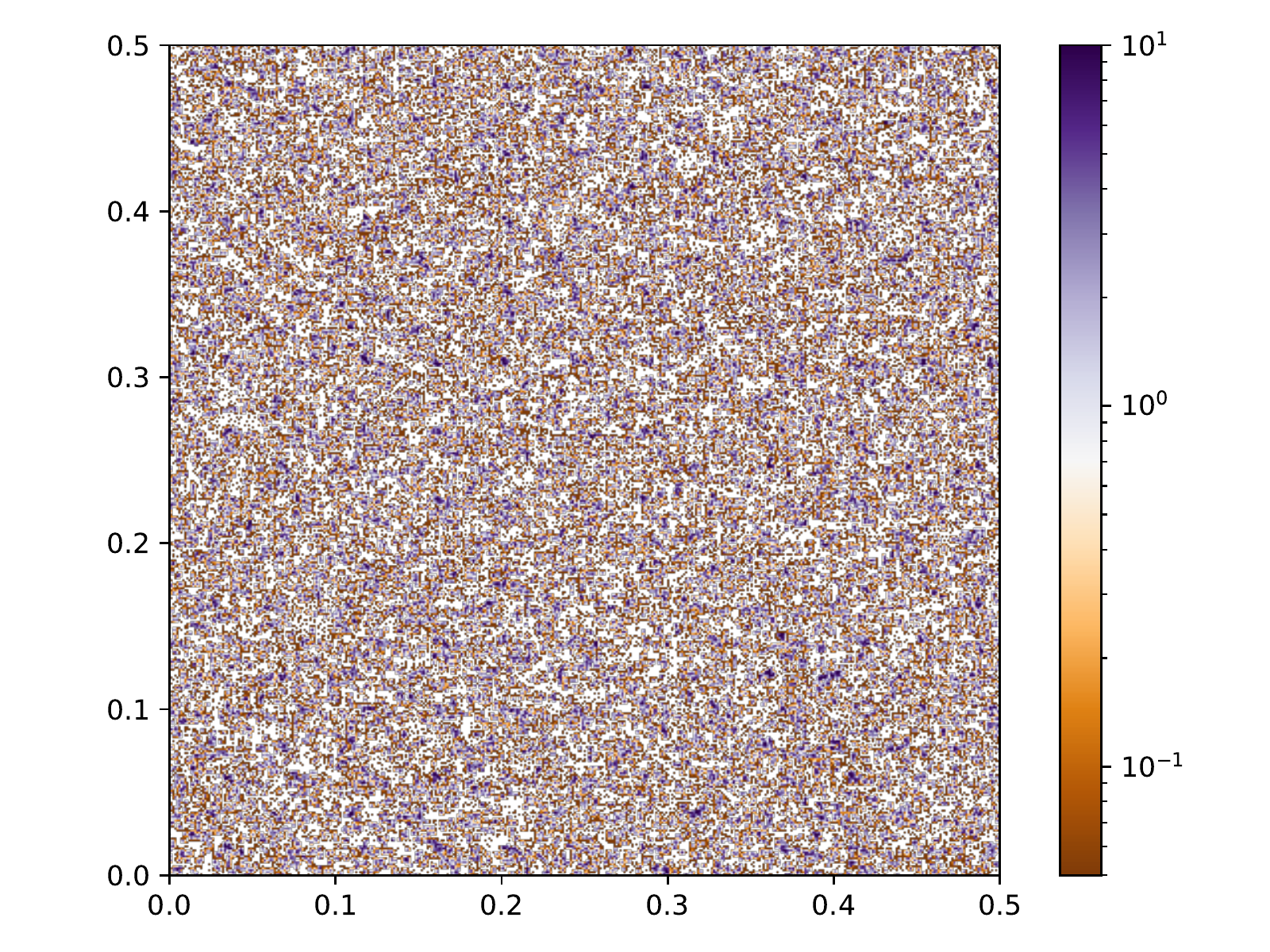}
  \plotone{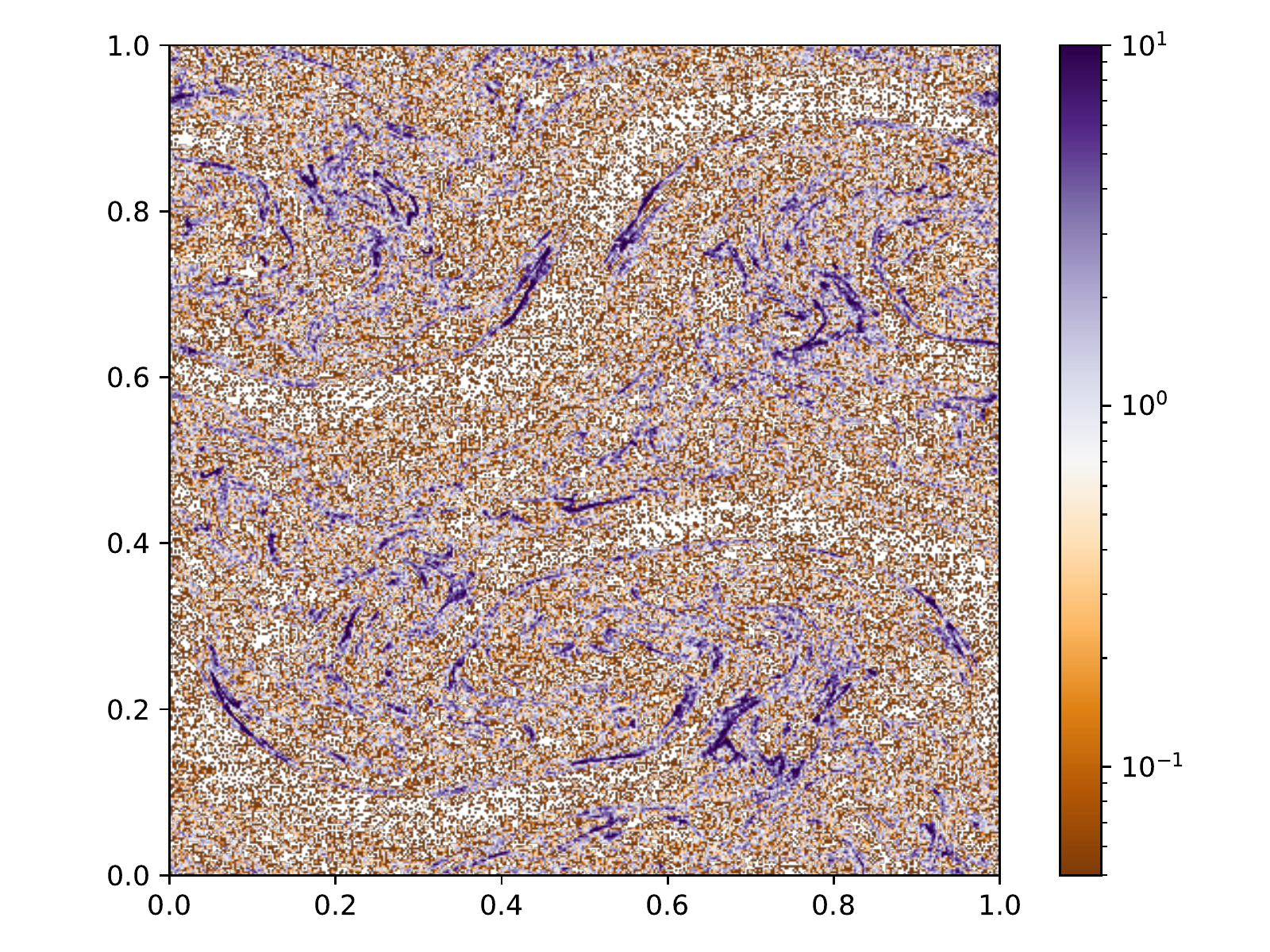}
  \caption{Dust-to-gas maps for single wave evolution at the
    final stage of the simulation. The upper
    panel corresponds to the shortest wavelength ($\lambda = 1/24$) and
    the simulation has been run in a domain half the size of the original
    to optimize the computational resources. 
     The lower panel
    corresponds to the longest wavelength
    ($\lambda = 1$). These plots show that the shortest wavelengths
    by themselves can only produce a clumpy distribution with
    localized overdensities, while the longest ones
    define the filamentary structures.
        \label{fig:compare_waves}
  }
\end{figure}

  Additionally, we studied separately
  the influence of the largest and shortest
  waves of the initial velocity spectrum in the
  formation of dust structures. As can be
  seen in Fig. \ref{fig:compare_waves},
  the shortest wave does not inject enough energy
    to the medium to produce conspicuous filamentary structures.
    Nevertheless, tiny clumps of a few ($\leq 10$) pixels are
    formed, and the histogram of the dust-to-gas ratio is
    very similar to the one for the complete simulation shown
    in Fig. \ref{fig:dgratio_histogram}. For the longest wave, the
    energy input is so strong that relevant fluctuations are
    appreciable in density, momentum, and magnetic field distributions. These
    variations enable the formation of dust filaments with
    a dust-to-gas ratio distribution that again resembles to
    the one in  Fig. \ref{fig:dgratio_histogram}.
    We draw two conclusions from this analysis: first, that variations
    in the dust-to-gas ratio are expected to arise whenever a wave
    propagates in a molecular cloud envelope, no matter the energy
    of the wave; and second, that the morphology of the structures
    that can be formed in such a medium depends on the wave spectrum.
    It is of interest to explore in a future work if other properties
    of the dust filaments, such as their widths, depend on the
    wave spectrum or if they have a characteristic size.
  Finally, we want to highlight again that the choice of the
    box size sets an upper limit to the longest wavelength that we are
    able to resolve ($L_{0}$), so we are always talking about stuctures
    at scales lower than $L_{0} = 1$ pc.  \par

\subsection{Limitations of the model}\label{subsec:limitations}

  There are some points to be taken into account when comparing these
  results with observations. Some of them are related to the necessary
  simplifications  incorporated in the current model,
  and others are linked to the typical plasma conditions
  of the MC envelopes that we have modelled.\par

The first point is that our model is 2D. This
implies that the motion of the particles is produced by polarized Alfvén waves. In 3D, the stirring of the plasma will be in 3D and the filaments too. Adding a degree of freedom will likely result in shorter, helical filaments..\par

The second point is that we have only evolved a single
population of dust with fixed radius, while in the
diffuse ISM a realistic size distribution that
evolves with time should be considered. With that purpose,
we have developed an additional module for Athena to follow dust
coagulation and shattering that will be presented in
a forthcoming paper (Beitia-Antero et al. in prep.).\par

And finally, the relevance of resistive processes, and
ambipolar and Hall diffusion should be taken into account.
The impact of these effects will be studied in subsequent works. \par

It is also relevant to highlight that the modeled phase is
very diffuse, partially ionised, and the dust particles
are very small ($a_{d} = 0.05~\mu$m) so it is better traced by
UV and optical tracers rather than with infrared/submillimeter ones.
Observations of MC envelopes
with high sensitivity, wide field, UV imagers will provide invaluable
results for this research.

\section{Conclusions}\label{sec:conclusions}
In this work, we have
described a new module developed for
  Athena to deal with the interaction of charged
  dust particles in astronomical magnetized fluids.
  The code is applied to the study of dust dynamics
  in diffuse molecular cloud envelopes, and its
  performance is illustrated through the formation
  of dust filaments decoupled from the gas.
The model for a molecular cloud envelope considers a
density of $n = 10$ cm$^{-3}$, with a temperature
of $T = 6000$ K and an ionization fraction $\chi = 0.1$,
and
includes a single population of charged
dust silicate particles of radius $0.05~\mu$m. \par

We show that charged
dust grains follow the magnetic
field and form filamentary structures with a typical width
of 0.0035-0.0087 pc. Although the mean dust-to-gas
ratio is globally preserved, local variations are observed
that are not always correlated with gas density;
these variations are in concordance with previous theoretical
\citep{2006ApJ...649..807P,2014ApJ...797...59H}
and observational \citep{2017ques.workE..27S,2015MNRAS.448.2187C,2015ApJ...811..118R} results. \par

Although a more realistic model (non-ideal MHD
effects and
a full dust distribution) is needed,
our results highlight the importance of
considering the dynamics of charged dust
for star formation
studies even at a low-density regime
characteristic of a molecular cloud envelope.

\acknowledgments

\textbf{We want to thank an anonymous referee for many helpful comments
  and suggestions that have helped to improve this manuscript.}
L. B.-A. acknowledges Universidad Complutense de Madrid and Banco
Santander for the grant ``Personal Investigador en
Formaci\'on CT17/17-CT18/17''. This work has been partially funded by the
Ministry of Economy and Competitiviness of Spain
through grants MINECO-ESP2015-68908-R and MINECO-ESP2017-87813-R.
L. B.-A. also wants to
acknowledge I. Prada for useful discussion about how to build the
dust maps and with the usage of the python image analysis packages. This
research
has made use of NASA’s Astrophysics Data System.

\appendix

\section{Code tests} \label{appendix:code_test}
We designed two very simple tests to check that the
movement of charged particles is correctly solved in 2D. Both tests
consider a positively charged grain, initially at rest,
embedded in a uniform medium moving at constant speed parallel to the
x-axis. The constant magnetic field is parallel to the z-axis, perpendicular
to the domain we are solving.

\subsection{Lorentz force term}
The first test is intended to check the integration accuracy
of the Lorentz force term. The problem to be solved analytically is:

\begin{eqnarray}
  \frac{d {\bf x}}{d t} = {\bf v}, ~~ {\bf x}(0) = (x_{0},x_{1}) \nonumber \\
  \frac{d {\bf v}}{d t} =
  - \frac{{\bf v} - {\bf u}}{t_{s}} + Q({\bf v} - {\bf u}) \times {\bf B},
  ~~{\bf v}(0) = (0,0) \nonumber
\end{eqnarray}

\begin{figure}[h!]
  \plotone{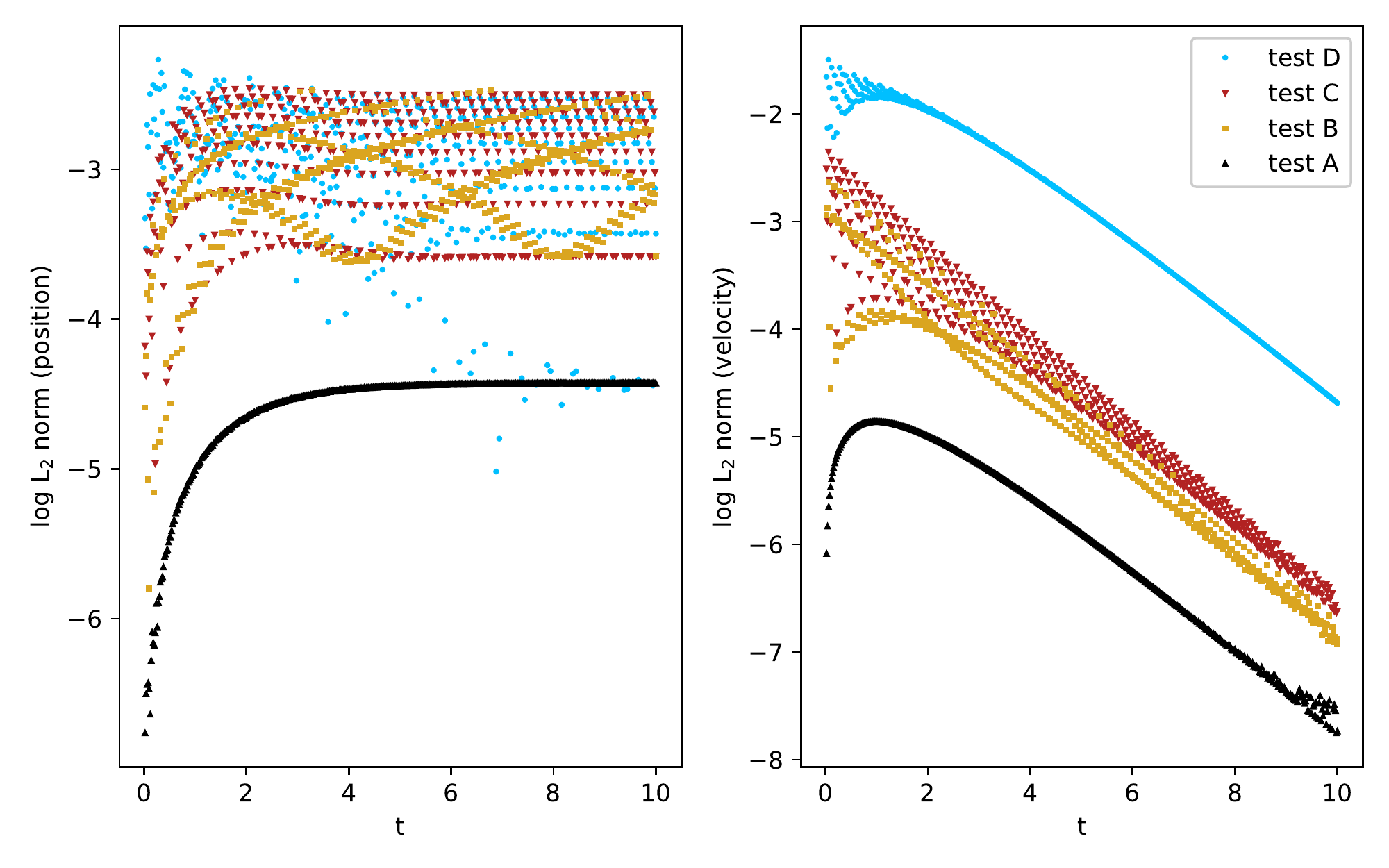}
  \caption{L2 norm for position (left) and velocity (right) for the test
    of the Lorentz force term and a CFL = 0.8.
    Black triangles correspond to test A, for
    which the aerodynamic drag is dominant. For tests B (orange squares),
    C (red updown triangles), and D (blue circles), the magnetic
    effects become increasingly dominant and oscillations in the
    errors are appreciable; however, they are always of the order of
    $10^{-3}$ or better for position, and $10^{-2}$ or better for velocity.
  }
  \label{fig:lorentz_test}
\end{figure}

Depending on the ratio
between $t_{s}$ and $Q|\mathbf{B}| \doteq QB$, the particle motion will be
dominated by gas drag ($t_{s}^{-1} \gg QB$) or by magnetic
effects ($t_{s}^{-1} \ll QB$). Hence, we run four tests
up to $t = 10t_{s}$ for a fiducial stopping time $t_{s} = 1$ and
have considered
the following values for the pair $(Q, B)$: test A (0.1, 0.1), test B
(0.1, 1), test C (0.1, 10), and test D (1, 10). For all the cases
(see Fig. \ref{fig:lorentz_test}) the error in position is of the order
of $10^{-3}$ or lower, and the error in velocity is at most of the
order of $10^{-2}$, which confirms that our integrator for the
Lorentz term works well and preserves the second-order accuracy. We want
to note that for a magnetic-dominated medium ($QB \sim 100$) it is
necessary to diminish even more the timestep in Eq. \ref{eq:modif_timestep} because
with the current implementation, the integration diverges after a few
timesteps. However, for studies of dust in the interstellar medium
such harsh conditions are not expected, so the implementation of the
integrator presented here is adequate for our purposes.

\subsection{Coulomb drag}
\begin{figure}[h!]
  \plotone{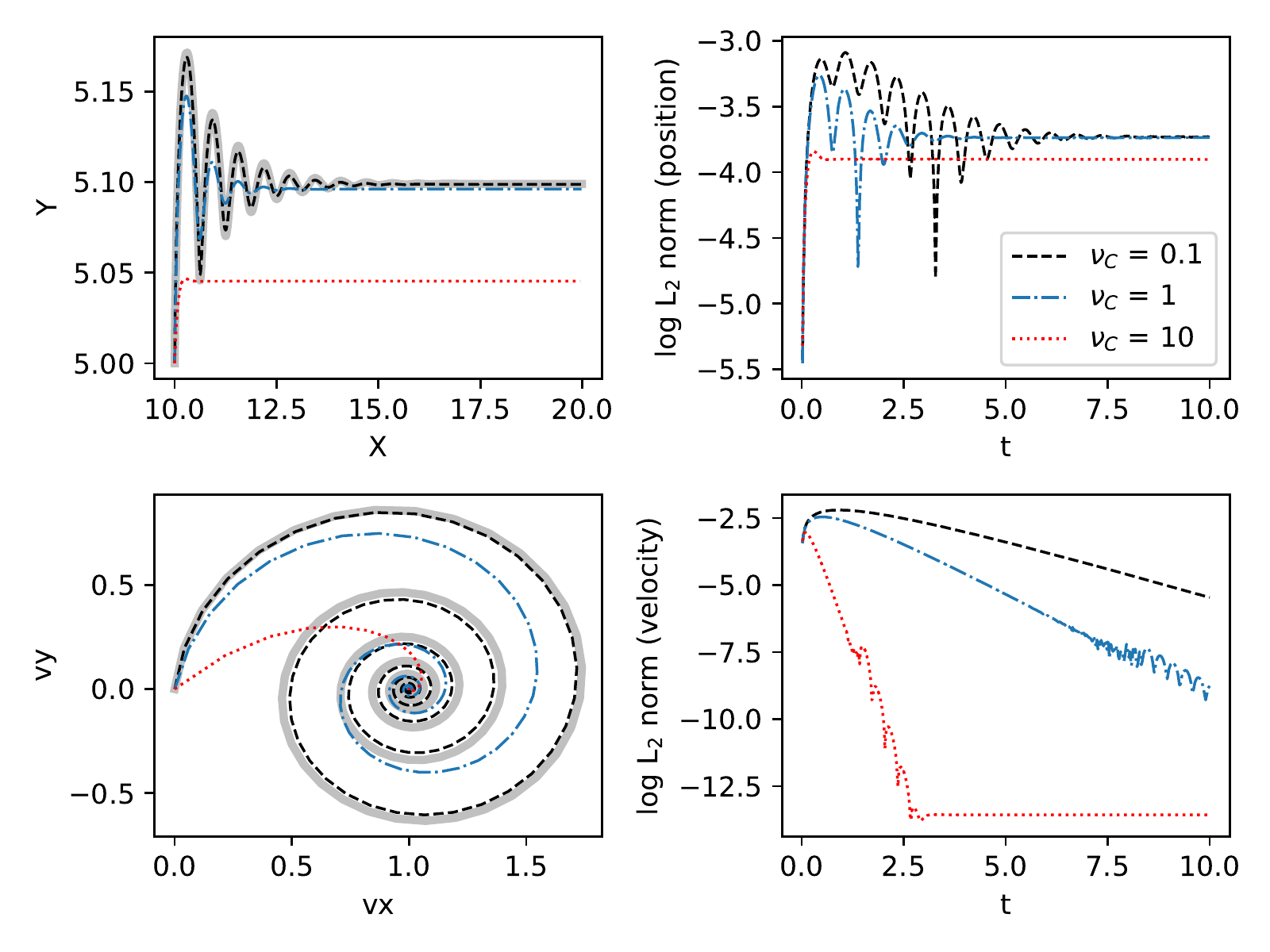}
  \caption{Results for the test of the Coulomb electric drag of a positively
    charged grain in the presence of a gas moving uniformly in the x-direction
    and a magnetic field oriented parallel to the z-axis. Position (upper
    left) and velocity (lower left) diagrams are shown for
    $\nu = 0.1$ (black dashed line), $\nu = 1$ (blue dashed-dotted line),
    and $\nu = 10$ (red dotted line); the thick light gray curve
    shows the dynamics of the charged particle without considering the
    Coulomb drag. $L_{2}$ norm of the errors
    in position (upper right) and velocity (lower right) are also shown.
  }
  \label{fig:coulomb_test}
\end{figure}

This test is defined to test the Coulomb drag term in the full
Eq. \ref{eq:dust_eom}. We run test D ($Q = 1$, $B = 10$)
for $\nu = 0.1, 1, 10$
to study how it effects the transition from the aerodynamic regime
to the magnetic one; the results are shown in Fig. \ref{fig:coulomb_test}
The main effect of the electric drag is the damping of the
magnetic oscillation of the charged grain, that can be rapidly
supressed for large values of $\nu$; however, it will be typically
lower than $Q$ since $\nu \propto Q^{2}$
(Eq. \ref{eq:coulomb_rate}). This implies
that the overall accuracy of the integrator is kept second-order, since
the oscillations in the precision arise from the effects of the Lorentz force.

\bibliography{references}{}

\begin{thebibliography}{}
\expandafter\ifx\csname natexlab\endcsname\relax\def\natexlab#1{#1}\fi
\providecommand{\url}[1]{\href{#1}{#1}}
\providecommand{\dodoi}[1]{doi:~\href{http://doi.org/#1}{\nolinkurl{#1}}}
\providecommand{\doeprint}[1]{\href{http://ascl.net/#1}{\nolinkurl{http://ascl.net/#1}}}
\providecommand{\doarXiv}[1]{\href{https://arxiv.org/abs/#1}{\nolinkurl{https://arxiv.org/abs/#1}}}

\bibitem[{{Andr{\'e}} {et~al.}(2010){Andr{\'e}}, {Men'shchikov}, {Bontemps},
  {K{\"o}nyves}, {Motte}, {Schneider}, {Didelon}, {Minier}, {Saraceno},
  {Ward-Thompson}, {di Francesco}, {White}, {Molinari}, {Testi}, {Abergel},
  {Griffin}, {Henning}, {Royer}, {Mer{\'\i}n}, {Vavrek}, {Attard},
  {Arzoumanian}, {Wilson}, {Ade}, {Aussel}, {Baluteau}, {Benedettini},
  {Bernard}, {Blommaert}, {Cambr{\'e}sy}, {Cox}, {di Giorgio}, {Hargrave},
  {Hennemann}, {Huang}, {Kirk}, {Krause}, {Launhardt}, {Leeks}, {Le Pennec},
  {Li}, {Martin}, {Maury}, {Olofsson}, {Omont}, {Peretto}, {Pezzuto}, {Prusti},
  {Roussel}, {Russeil}, {Sauvage}, {Sibthorpe}, {Sicilia-Aguilar}, {Spinoglio},
  {Waelkens}, {Woodcraft}, \& {Zavagno}}]{2010A&A...518L.102A}
{Andr{\'e}}, P., {Men'shchikov}, A., {Bontemps}, S., {et~al.} 2010, \aap, 518,
  L102, \dodoi{10.1051/0004-6361/201014666}

\bibitem[{{Arzoumanian} {et~al.}(2011){Arzoumanian}, {Andr{\'e}}, {Didelon},
  {K{\"o}nyves}, {Schneider}, {Men'shchikov}, {Sousbie}, {Zavagno}, {Bontemps},
  {di Francesco}, {Griffin}, {Hennemann}, {Hill}, {Kirk}, {Martin}, {Minier},
  {Molinari}, {Motte}, {Peretto}, {Pezzuto}, {Spinoglio}, {Ward-Thompson},
  {White}, \& {Wilson}}]{2011A&A...529L...6A}
{Arzoumanian}, D., {Andr{\'e}}, P., {Didelon}, P., {et~al.} 2011, \aap, 529,
  L6, \dodoi{10.1051/0004-6361/201116596}

\bibitem[{{Bacchini} {et~al.}(2020){Bacchini}, {Fraternali}, {Iorio},
  {Pezzulli}, {Marasco}, \& {Nipoti}}]{2020arXiv200610764B}
{Bacchini}, C., {Fraternali}, F., {Iorio}, G., {et~al.} 2020, arXiv e-prints,
  arXiv:2006.10764.
\newblock \doarXiv{2006.10764}

\bibitem[{{Bai} \& {Stone}(2010)}]{2010ApJS..190..297B}
{Bai}, X.-N., \& {Stone}, J.~M. 2010, \apjs, 190, 297,
  \dodoi{10.1088/0067-0049/190/2/297}

\bibitem[{{Bakes} \& {Tielens}(1994)}]{1994ApJ...427..822B}
{Bakes}, E.~L.~O., \& {Tielens}, A.~G.~G.~M. 1994, \apj, 427, 822,
  \dodoi{10.1086/174188}

\bibitem[{{Ballesteros-Paredes} {et~al.}(2020){Ballesteros-Paredes},
  {Andr{\'e}}, {Hennebelle}, {Klessen}, {Kruijssen}, {Chevance}, {Nakamura},
  {Adamo}, \& {V{\'a}zquez-Semadeni}}]{2020SSRv..216...76B}
{Ballesteros-Paredes}, J., {Andr{\'e}}, P., {Hennebelle}, P., {et~al.} 2020,
  \ssr, 216, 76, \dodoi{10.1007/s11214-020-00698-3}

\bibitem[{{Balsara}(1996)}]{1996ApJ...465..775B}
{Balsara}, D.~S. 1996, \apj, 465, 775, \dodoi{10.1086/177462}

\bibitem[{{Beitia-Antero} \& {G{\'o}mez de Castro}(2017)}]{2017MNRAS.469.2531B}
{Beitia-Antero}, L., \& {G{\'o}mez de Castro}, A.~I. 2017, \mnras, 469, 2531,
  \dodoi{10.1093/mnras/stx881}

\bibitem[{{Blitz} {et~al.}(2007){Blitz}, {Fukui}, {Kawamura}, {Leroy},
  {Mizuno}, \& {Rosolowsky}}]{2007prpl.conf...81B}
{Blitz}, L., {Fukui}, Y., {Kawamura}, A., {et~al.} 2007, in Protostars and
  Planets V, ed. B.~{Reipurth}, D.~{Jewitt}, \& K.~{Keil}, 81.
\newblock \doarXiv{astro-ph/0602600}

\bibitem[{{Bohlin} {et~al.}(1978){Bohlin}, {Savage}, \&
  {Drake}}]{1978ApJ...224..132B}
{Bohlin}, R.~C., {Savage}, B.~D., \& {Drake}, J.~F. 1978, \apj, 224, 132,
  \dodoi{10.1086/156357}

\bibitem[{Boris(1970)}]{Boris1970}
Boris, J. 1970, in Proc. Fourth Conf. on Numerical Simulation of Plasmas
  (Washington, DC: Naval Research Lab), 3

\bibitem[{{Boulanger} {et~al.}(1996){Boulanger}, {Abergel}, {Bernard},
  {Burton}, {Desert}, {Hartmann}, {Lagache}, \& {Puget}}]{1996A&A...312..256B}
{Boulanger}, F., {Abergel}, A., {Bernard}, J.~P., {et~al.} 1996, \aap, 312, 256

\bibitem[{{Cazaux} \& {Tielens}(2004)}]{2004ApJ...604..222C}
{Cazaux}, S., \& {Tielens}, A.~G.~G.~M. 2004, \apj, 604, 222,
  \dodoi{10.1086/381775}

\bibitem[{{Chapman} \& {Wardle}(2006)}]{2006MNRAS.371..513C}
{Chapman}, J.~F., \& {Wardle}, M. 2006, \mnras, 371, 513,
  \dodoi{10.1111/j.1365-2966.2006.10592.x}

\bibitem[{{Chen} {et~al.}(2015){Chen}, {Liu}, {Yuan}, {Huang}, \&
  {Xiang}}]{2015MNRAS.448.2187C}
{Chen}, B.~Q., {Liu}, X.~W., {Yuan}, H.~B., {Huang}, Y., \& {Xiang}, M.~S.
  2015, \mnras, 448, 2187, \dodoi{10.1093/mnras/stv103}

\bibitem[{{Chevance} {et~al.}(2020){Chevance}, {Kruijssen}, {Vazquez-Semadeni},
  {Nakamura}, {Klessen}, {Ballesteros-Paredes}, {Inutsuka}, {Adamo}, \&
  {Hennebelle}}]{2020SSRv..216...50C}
{Chevance}, M., {Kruijssen}, J.~M.~D., {Vazquez-Semadeni}, E., {et~al.} 2020,
  \ssr, 216, 50, \dodoi{10.1007/s11214-020-00674-x}

\bibitem[{{Clark} {et~al.}(2014){Clark}, {Peek}, \&
  {Putman}}]{2014ApJ...789...82C}
{Clark}, S.~E., {Peek}, J.~E.~G., \& {Putman}, M.~E. 2014, \apj, 789, 82,
  \dodoi{10.1088/0004-637X/789/1/82}

\bibitem[{{Cramer} \& {Vladimirov}(1997)}]{1997PASA...14..170C}
{Cramer}, N.~F., \& {Vladimirov}, S.~V. 1997, \pasa, 14, 170,
  \dodoi{10.1071/AS97170}

\bibitem[{{Elmegreen} \& {Scalo}(2004)}]{2004ARA&A..42..211E}
{Elmegreen}, B.~G., \& {Scalo}, J. 2004, \araa, 42, 211,
  \dodoi{10.1146/annurev.astro.41.011802.094859}

\bibitem[{{Falceta-Gon{\c{c}}alves} {et~al.}(2003){Falceta-Gon{\c{c}}alves},
  {de Juli}, \& {Jatenco-Pereira}}]{2003ApJ...597..970F}
{Falceta-Gon{\c{c}}alves}, D., {de Juli}, M.~C., \& {Jatenco-Pereira}, V. 2003,
  \apj, 597, 970, \dodoi{10.1086/378584}

\bibitem[{{Falgarone} \& {Phillips}(1996)}]{1996ApJ...472..191F}
{Falgarone}, E., \& {Phillips}, T.~G. 1996, \apj, 472, 191,
  \dodoi{10.1086/178054}

\bibitem[{{Falgarone} {et~al.}(2005){Falgarone}, {Verstraete}, {Pineau Des
  For{\^e}ts}, \& {Hily-Blant}}]{2005A&A...433..997F}
{Falgarone}, E., {Verstraete}, L., {Pineau Des For{\^e}ts}, G., \&
  {Hily-Blant}, P. 2005, \aap, 433, 997, \dodoi{10.1051/0004-6361:20041893}

\bibitem[{{Federrath}(2015)}]{2015MNRAS.450.4035F}
{Federrath}, C. 2015, \mnras, 450, 4035, \dodoi{10.1093/mnras/stv941}

\bibitem[{{Ferri{\`e}re}(2001)}]{2001RvMP...73.1031F}
{Ferri{\`e}re}, K.~M. 2001, Reviews of Modern Physics, 73, 1031,
  \dodoi{10.1103/RevModPhys.73.1031}

\bibitem[{{Fiege} \& {Pudritz}(2000)}]{2000MNRAS.311..105F}
{Fiege}, J.~D., \& {Pudritz}, R.~E. 2000, \mnras, 311, 105,
  \dodoi{10.1046/j.1365-8711.2000.03067.x}

\bibitem[{{Fissel} {et~al.}(2019){Fissel}, {Ade}, {Angil{\`e}}, {Ashton},
  {Benton}, {Chen}, {Cunningham}, {Devlin}, {Dober}, {Friesen}, {Fukui},
  {Galitzki}, {Gandilo}, {Goodman}, {Green}, {Jones}, {Klein}, {King},
  {Korotkov}, {Li}, {Lowe}, {Martin}, {Matthews}, {Moncelsi}, {Nakamura},
  {Netterfield}, {Newmark}, {Novak}, {Pascale}, {Poidevin}, {Santos}, {Savini},
  {Scott}, {Shariff}, {Soler}, {Thomas}, {Tucker}, {Tucker}, {Ward-Thompson},
  \& {Zucker}}]{2019ApJ...878..110F}
{Fissel}, L.~M., {Ade}, P. A.~R., {Angil{\`e}}, F.~E., {et~al.} 2019, \apj,
  878, 110, \dodoi{10.3847/1538-4357/ab1eb0}

\bibitem[{{Flagey} {et~al.}(2009){Flagey}, {Noriega-Crespo}, {Boulanger},
  {Carey}, {Brooke}, {Falgarone}, {Huard}, {McCabe}, {Miville-Desch{\^e}nes},
  {Padgett}, {Paladini}, \& {Rebull}}]{2009ApJ...701.1450F}
{Flagey}, N., {Noriega-Crespo}, A., {Boulanger}, F., {et~al.} 2009, \apj, 701,
  1450, \dodoi{10.1088/0004-637X/701/2/1450}

\bibitem[{{Fukui} {et~al.}(2017){Fukui}, {Tsuge}, {Sano}, {Bekki}, {Yozin},
  {Tachihara}, \& {Inoue}}]{2017PASJ...69L...5F}
{Fukui}, Y., {Tsuge}, K., {Sano}, H., {et~al.} 2017, \pasj, 69, L5,
  \dodoi{10.1093/pasj/psx032}

\bibitem[{{Fukui} {et~al.}(2009){Fukui}, {Kawamura}, {Wong}, {Murai},
  {Iritani}, {Mizuno}, {Mizuno}, {Onishi}, {Hughes}, {Ott}, {Muller},
  {Staveley-Smith}, \& {Kim}}]{2009ApJ...705..144F}
{Fukui}, Y., {Kawamura}, A., {Wong}, T., {et~al.} 2009, \apj, 705, 144,
  \dodoi{10.1088/0004-637X/705/1/144}

\bibitem[{{G{\'o}mez de Castro} {et~al.}(2015){G{\'o}mez de Castro},
  {L{\'o}pez-Santiago}, {L{\'o}pez-Mart{\'\i}nez}, {S{\'a}nchez}, {de Castro},
  \& {Cornide}}]{2015MNRAS.449.3867G}
{G{\'o}mez de Castro}, A.~I., {L{\'o}pez-Santiago}, J.,
  {L{\'o}pez-Mart{\'\i}nez}, F., {et~al.} 2015, \mnras, 449, 3867,
  \dodoi{10.1093/mnras/stv413}

\bibitem[{{Hocuk} \& {Spaans}(2010)}]{2010A&A...510A.110H}
{Hocuk}, S., \& {Spaans}, M. 2010, \aap, 510, A110,
  \dodoi{10.1051/0004-6361/200912236}

\bibitem[{{Hollenbach} \& {Salpeter}(1971)}]{1971ApJ...163..155H}
{Hollenbach}, D., \& {Salpeter}, E.~E. 1971, \apj, 163, 155,
  \dodoi{10.1086/150754}

\bibitem[{{Hopkins}(2014)}]{2014ApJ...797...59H}
{Hopkins}, P.~F. 2014, \apj, 797, 59, \dodoi{10.1088/0004-637X/797/1/59}

\bibitem[{{Hopkins} \& {Squire}(2018)}]{2018MNRAS.479.4681H}
{Hopkins}, P.~F., \& {Squire}, J. 2018, \mnras, 479, 4681,
  \dodoi{10.1093/mnras/sty1604}

\bibitem[{{Hopkins} {et~al.}(2020){Hopkins}, {Squire}, \&
  {Seligman}}]{2020MNRAS.496.2123H}
{Hopkins}, P.~F., {Squire}, J., \& {Seligman}, D. 2020, \mnras, 496, 2123,
  \dodoi{10.1093/mnras/staa1046}

\bibitem[{{Inutsuka} {et~al.}(2015){Inutsuka}, {Inoue}, {Iwasaki}, \&
  {Hosokawa}}]{2015A&A...580A..49I}
{Inutsuka}, S.-i., {Inoue}, T., {Iwasaki}, K., \& {Hosokawa}, T. 2015, \aap,
  580, A49, \dodoi{10.1051/0004-6361/201425584}

\bibitem[{{Ivlev} {et~al.}(2018){Ivlev}, {Dogiel}, {Chernyshov}, {Caselli},
  {Ko}, \& {Cheng}}]{2018ApJ...855...23I}
{Ivlev}, A.~V., {Dogiel}, V.~A., {Chernyshov}, D.~O., {et~al.} 2018, \apj, 855,
  23, \dodoi{10.3847/1538-4357/aaadb9}

\bibitem[{{Ivlev} {et~al.}(2015){Ivlev}, {Padovani}, {Galli}, \&
  {Caselli}}]{2015ApJ...812..135I}
{Ivlev}, A.~V., {Padovani}, M., {Galli}, D., \& {Caselli}, P. 2015, \apj, 812,
  135, \dodoi{10.1088/0004-637X/812/2/135}

\bibitem[{{Klessen} \& {Glover}(2016)}]{2016SAAS...43...85K}
{Klessen}, R.~S., \& {Glover}, S. C.~O. 2016, Saas-Fee Advanced Course, 43, 85,
  \dodoi{10.1007/978-3-662-47890-5_2}

\bibitem[{{Koch} \& {Rosolowsky}(2015)}]{2015MNRAS.452.3435K}
{Koch}, E.~W., \& {Rosolowsky}, E.~W. 2015, \mnras, 452, 3435,
  \dodoi{10.1093/mnras/stv1521}

\bibitem[{{K{\"o}hler} {et~al.}(2015){K{\"o}hler}, {Ysard}, \&
  {Jones}}]{2015A&A...579A..15K}
{K{\"o}hler}, M., {Ysard}, N., \& {Jones}, A.~P. 2015, \aap, 579, A15,
  \dodoi{10.1051/0004-6361/201525646}

\bibitem[{{Kruijssen} {et~al.}(2019){Kruijssen}, {Dale}, {Longmore}, {Walker},
  {Henshaw}, {Jeffreson}, {Petkova}, {Ginsburg}, {Barnes}, {Battersby},
  {Immer}, {Jackson}, {Keto}, {Krieger}, {Mills}, {S{\'a}nchez-Monge},
  {Schmiedeke}, {Suri}, \& {Zhang}}]{2019MNRAS.484.5734K}
{Kruijssen}, J.~M.~D., {Dale}, J.~E., {Longmore}, S.~N., {et~al.} 2019, \mnras,
  484, 5734, \dodoi{10.1093/mnras/stz381}

\bibitem[{{Lallement} {et~al.}(2003){Lallement}, {Welsh}, {Vergely}, {Crifo},
  \& {Sfeir}}]{2003A&A...411..447L}
{Lallement}, R., {Welsh}, B.~Y., {Vergely}, J.~L., {Crifo}, F., \& {Sfeir}, D.
  2003, \aap, 411, 447, \dodoi{10.1051/0004-6361:20031214}

\bibitem[{{Lazarian} \& {Yan}(2002)}]{2002ApJ...566L.105L}
{Lazarian}, A., \& {Yan}, H. 2002, \apjl, 566, L105, \dodoi{10.1086/339675}

\bibitem[{{Lee} {et~al.}(2017){Lee}, {Hopkins}, \&
  {Squire}}]{2017MNRAS.469.3532L}
{Lee}, H., {Hopkins}, P.~F., \& {Squire}, J. 2017, \mnras, 469, 3532,
  \dodoi{10.1093/mnras/stx1097}

\bibitem[{{Lehe} {et~al.}(2009){Lehe}, {Parrish}, \&
  {Quataert}}]{2009ApJ...707..404L}
{Lehe}, R., {Parrish}, I.~J., \& {Quataert}, E. 2009, \apj, 707, 404,
  \dodoi{10.1088/0004-637X/707/1/404}

\bibitem[{{Lenz} {et~al.}(2017){Lenz}, {Hensley}, \&
  {Dor{\'e}}}]{2017ApJ...846...38L}
{Lenz}, D., {Hensley}, B.~S., \& {Dor{\'e}}, O. 2017, \apj, 846, 38,
  \dodoi{10.3847/1538-4357/aa84af}

\bibitem[{{Martin} {et~al.}(2015){Martin}, {Blagrave}, {Lockman}, {Pinheiro
  Gon{\c{c}}alves}, {Boothroyd}, {Joncas}, {Miville-Desch{\^e}nes}, \&
  {Stephan}}]{2015ApJ...809..153M}
{Martin}, P.~G., {Blagrave}, K.~P.~M., {Lockman}, F.~J., {et~al.} 2015, \apj,
  809, 153, \dodoi{10.1088/0004-637X/809/2/153}

\bibitem[{{Mathis} {et~al.}(1983){Mathis}, {Mezger}, \&
  {Panagia}}]{1983A&A...128..212M}
{Mathis}, J.~S., {Mezger}, P.~G., \& {Panagia}, N. 1983, \aap, 500, 259

\bibitem[{{Meidt} {et~al.}(2018){Meidt}, {Leroy}, {Rosolowsky}, {Kruijssen},
  {Schinnerer}, {Schruba}, {Pety}, {Blanc}, {Bigiel}, {Chevance}, {Hughes},
  {Querejeta}, \& {Usero}}]{2018ApJ...854..100M}
{Meidt}, S.~E., {Leroy}, A.~K., {Rosolowsky}, E., {et~al.} 2018, \apj, 854,
  100, \dodoi{10.3847/1538-4357/aaa290}

\bibitem[{{Mignone} {et~al.}(2019){Mignone}, {Flock}, \&
  {Vaidya}}]{2019ApJS..244...38M}
{Mignone}, A., {Flock}, M., \& {Vaidya}, B. 2019, \apjs, 244, 38,
  \dodoi{10.3847/1538-4365/ab4356}

\bibitem[{{Myers} \& {Goodman}(1988)}]{1988ApJ...329..392M}
{Myers}, P.~C., \& {Goodman}, A.~A. 1988, \apj, 329, 392,
  \dodoi{10.1086/166385}

\bibitem[{{Nakamura} \& {Li}(2008)}]{2008ApJ...687..354N}
{Nakamura}, F., \& {Li}, Z.-Y. 2008, \apj, 687, 354, \dodoi{10.1086/591641}

\bibitem[{{Nakano}(1998)}]{1998ApJ...494..587N}
{Nakano}, T. 1998, \apj, 494, 587, \dodoi{10.1086/305230}

\bibitem[{{Offner} \& {Liu}(2018)}]{2018NatAs...2..896O}
{Offner}, S. S.~R., \& {Liu}, Y. 2018, Nature Astronomy, 2, 896,
  \dodoi{10.1038/s41550-018-0566-1}

\bibitem[{{Padoan} {et~al.}(2006){Padoan}, {Cambr{\'e}sy}, {Juvela}, {Kritsuk},
  {Langer}, \& {Norman}}]{2006ApJ...649..807P}
{Padoan}, P., {Cambr{\'e}sy}, L., {Juvela}, M., {et~al.} 2006, \apj, 649, 807,
  \dodoi{10.1086/507068}

\bibitem[{{Padoan} {et~al.}(2016){Padoan}, {Pan}, {Haugb{\o}lle}, \&
  {Nordlund}}]{2016ApJ...822...11P}
{Padoan}, P., {Pan}, L., {Haugb{\o}lle}, T., \& {Nordlund}, {\r{A}}. 2016,
  \apj, 822, 11, \dodoi{10.3847/0004-637X/822/1/11}

\bibitem[{{Palmeirim} {et~al.}(2013){Palmeirim}, {Andr{\'e}}, {Kirk},
  {Ward-Thompson}, {Arzoumanian}, {K{\"o}nyves}, {Didelon}, {Schneider},
  {Benedettini}, {Bontemps}, {Di Francesco}, {Elia}, {Griffin}, {Hennemann},
  {Hill}, {Martin}, {Men'shchikov}, {Molinari}, {Motte}, {Nguyen Luong},
  {Nutter}, {Peretto}, {Pezzuto}, {Roy}, {Rygl}, {Spinoglio}, \&
  {White}}]{2013A&A...550A..38P}
{Palmeirim}, P., {Andr{\'e}}, P., {Kirk}, J., {et~al.} 2013, \aap, 550, A38,
  \dodoi{10.1051/0004-6361/201220500}

\bibitem[{{Pandey} \& {Vladimirov}(2019)}]{2019AJ....157...83P}
{Pandey}, B.~P., \& {Vladimirov}, S.~V. 2019, \aj, 157, 83,
  \dodoi{10.3847/1538-3881/aafc32}

\bibitem[{{Panopoulou} {et~al.}(2017){Panopoulou}, {Psaradaki}, {Skalidis},
  {Tassis}, \& {Andrews}}]{2017MNRAS.466.2529P}
{Panopoulou}, G.~V., {Psaradaki}, I., {Skalidis}, R., {Tassis}, K., \&
  {Andrews}, J.~J. 2017, \mnras, 466, 2529, \dodoi{10.1093/mnras/stw3060}

\bibitem[{Pedregosa {et~al.}(2011)Pedregosa, Varoquaux, Gramfort, Michel,
  Thirion, Grisel, Blondel, Prettenhofer, Weiss, Dubourg, Vanderplas, Passos,
  Cournapeau, Brucher, Perrot, \& Duchesnay}]{scikit-learn}
Pedregosa, F., Varoquaux, G., Gramfort, A., {et~al.} 2011, Journal of Machine
  Learning Research, 12, 2825

\bibitem[{{Pilipp} {et~al.}(1987){Pilipp}, {Hartquist}, {Havnes}, \&
  {Morfill}}]{1987ApJ...314..341P}
{Pilipp}, W., {Hartquist}, T.~W., {Havnes}, O., \& {Morfill}, G.~E. 1987, \apj,
  314, 341, \dodoi{10.1086/165064}

\bibitem[{{Pinto} {et~al.}(2008){Pinto}, {Galli}, \&
  {Bacciotti}}]{2008A&A...484....1P}
{Pinto}, C., {Galli}, D., \& {Bacciotti}, F. 2008, \aap, 484, 1,
  \dodoi{10.1051/0004-6361:20078818}

\bibitem[{{Planck Collaboration} {et~al.}(2016){Planck Collaboration}, {Adam},
  {Ade}, {Aghanim}, {Alves}, {Arnaud}, {Arzoumanian}, {Ashdown}, {Aumont},
  {Baccigalupi}, {Banday}, {Barreiro}, {Bartolo}, {Battaner}, {Benabed},
  {Benoit-L{\'e}vy}, {Bernard}, {Bersanelli}, {Bielewicz}, {Bonaldi},
  {Bonavera}, {Bond}, {Borrill}, {Bouchet}, {Boulanger}, {Bracco}, {Burigana},
  {Butler}, {Calabrese}, {Cardoso}, {Catalano}, {Chamballu}, {Chiang},
  {Christensen}, {Colombi}, {Colombo}, {Combet}, {Couchot}, {Crill}, {Curto},
  {Cuttaia}, {Danese}, {Davies}, {Davis}, {de Bernardis}, {de Rosa}, {de
  Zotti}, {Delabrouille}, {Dickinson}, {Diego}, {Dole}, {Donzelli}, {Dor{\'e}},
  {Douspis}, {Ducout}, {Dupac}, {Efstathiou}, {Elsner}, {En{\ss}lin},
  {Eriksen}, {Falgarone}, {Ferri{\`e}re}, {Finelli}, {Forni}, {Frailis},
  {Fraisse}, {Franceschi}, {Frejsel}, {Galeotta}, {Galli}, {Ganga}, {Ghosh},
  {Giard}, {Gjerl{\o}w}, {Gonz{\'a}lez-Nuevo}, {G{\'o}rski}, {Gregorio},
  {Gruppuso}, {Guillet}, {Hansen}, {Hanson}, {Harrison}, {Henrot-Versill{\'e}},
  {Hern{\'a}ndez-Monteagudo}, {Herranz}, {Hildebrand t}, {Hivon}, {Hobson},
  {Holmes}, {Hovest}, {Huffenberger}, {Hurier}, {Jaffe}, {Jaffe}, {Jones},
  {Juvela}, {Keih{\"a}nen}, {Keskitalo}, {Kisner}, {Kneissl}, {Knoche}, {Kunz},
  {Kurki-Suonio}, {Lagache}, {Lamarre}, {Lasenby}, {Lattanzi}, {Lawrence},
  {Leonardi}, {Levrier}, {Liguori}, {Lilje}, {Linden-V{\o}rnle},
  {L{\'o}pez-Caniego}, {Lubin}, {Mac{\'\i}as-P{\'e}rez}, {Maffei}, {Maino},
  {Mand olesi}, {Maris}, {Marshall}, {Martin}, {Mart{\'\i}nez-Gonz{\'a}lez},
  {Masi}, {Matarrese}, {Mazzotta}, {Melchiorri}, {Mendes}, {Mennella},
  {Migliaccio}, {Miville-Desch{\^e}nes}, {Moneti}, {Montier}, {Morgante},
  {Mortlock}, {Munshi}, {Murphy}, {Naselsky}, {Natoli}, {N{\o}rgaard-Nielsen},
  {Noviello}, {Novikov}, {Novikov}, {Oppermann}, {Oxborrow}, {Pagano}, {Pajot},
  {Paoletti}, {Pasian}, {Perdereau}, {Perotto}, {Perrotta}, {Pettorino},
  {Piacentini}, {Piat}, {Plaszczynski}, {Pointecouteau}, {Polenta}, {Ponthieu},
  {Popa}, {Pratt}, {Prunet}, {Puget}, {Rachen}, {Reach}, {Reinecke},
  {Remazeilles}, {Renault}, {Ristorcelli}, {Rocha}, {Roudier},
  {Rubi{\~n}o-Mart{\'\i}n}, {Rusholme}, {Sandri}, {Santos}, {Savini}, {Scott},
  {Soler}, {Spencer}, {Stolyarov}, {Sudiwala}, {Sunyaev}, {Sutton},
  {Suur-Uski}, {Sygnet}, {Tauber}, {Terenzi}, {Toffolatti}, {Tomasi},
  {Tristram}, {Tucci}, {Umana}, {Valenziano}, {Valiviita}, {Van Tent},
  {Vielva}, {Villa}, {Wade}, {Wandelt}, {Wehus}, {Wiesemeyer}, {Yvon},
  {Zacchei}, \& {Zonca}}]{2016A&A...586A.135P}
{Planck Collaboration}, {Adam}, R., {Ade}, P.~A.~R., {et~al.} 2016, \aap, 586,
  A135, \dodoi{10.1051/0004-6361/201425044}

\bibitem[{{Planck Collaboration} {et~al.}(2018{\natexlab{a}}){Planck
  Collaboration}, {Akrami}, {Ashdown}, {Aumont}, {Baccigalupi}, {Ballardini},
  {Band ay}, {Barreiro}, {Bartolo}, {Basak}, {Benabed}, {Bernard},
  {Bersanelli}, {Bielewicz}, {Bond}, {Borrill}, {Bouchet}, {Boulanger},
  {Bracco}, {Bucher}, {Burigana}, {Calabrese}, {Cardoso}, {Carron}, {Chiang},
  {Combet}, {Crill}, {de Bernardis}, {de Zotti}, {Delabrouille}, {Delouis}, {Di
  Valentino}, {Dickinson}, {Diego}, {Ducout}, {Dupac}, {Efstathiou}, {Elsner},
  {En{\ss}lin}, {Falgarone}, {Fantaye}, {Ferri{\`e}re}, {Finelli},
  {Forastieri}, {Frailis}, {Fraisse}, {Franceschi}, {Frolov}, {Galeotta},
  {Galli}, {Ganga}, {G{\'e}nova-Santos}, {Ghosh}, {Gonz{\'a}lez-Nuevo},
  {G{\'o}rski}, {Gruppuso}, {Gudmundsson}, {Guillet}, {Handley}, {Hansen},
  {Herranz}, {Huang}, {Jaffe}, {Jones}, {Keih{\"a}nen}, {Keskitalo}, {Kiiveri},
  {Kim}, {Krachmalnicoff}, {Kunz}, {Kurki-Suonio}, {Lamarre}, {Lasenby}, {Le
  Jeune}, {Levrier}, {Liguori}, {Lilje}, {Lindholm}, {L{\'o}pez-Caniego},
  {Lubin}, {Ma}, {Mac{\'\i}as-P{\'e}rez}, {Maggio}, {Maino}, {Mandolesi},
  {Mangilli}, {Martin}, {Mart{\'\i}nez-Gonz{\'a}lez}, {Matarrese}, {McEwen},
  {Meinhold}, {Melchiorri}, {Migliaccio}, {Miville-Desch{\^e}nes}, {Molinari},
  {Moneti}, {Montier}, {Morgante}, {Natoli}, {Pagano}, {Paoletti}, {Pettorino},
  {Piacentini}, {Polenta}, {Puget}, {Rachen}, {Reinecke}, {Remazeilles},
  {Renzi}, {Rocha}, {Rosset}, {Roudier}, {Rubi{\~n}o-Mart{\'\i}n},
  {Ruiz-Granados}, {Salvati}, {Sandri}, {Savelainen}, {Scott}, {Soler},
  {Spencer}, {Tauber}, {Tavagnacco}, {Toffolatti}, {Tomasi}, {Trombetti},
  {Valiviita}, {Vansyngel}, {Van Tent}, {Vielva}, {Villa}, {Vittorio}, {Wehus},
  {Zacchei}, \& {Zonca}}]{2018arXiv180104945P}
{Planck Collaboration}, {Akrami}, Y., {Ashdown}, M., {et~al.}
  2018{\natexlab{a}}, arXiv e-prints, arXiv:1801.04945.
\newblock \doarXiv{1801.04945}

\bibitem[{{Planck Collaboration} {et~al.}(2018{\natexlab{b}}){Planck
  Collaboration}, {Aghanim}, {Akrami}, {Alves}, {Ashdown}, {Aumont},
  {Baccigalupi}, {Ballardini}, {Banday}, {Barreiro}, {Bartolo}, {Basak},
  {Benabed}, {Bernard}, {Bersanelli}, {Bielewicz}, {Bock}, {Bond}, {Borrill},
  {Bouchet}, {Boulanger}, {Bracco}, {Bucher}, {Burigana}, {Calabrese},
  {Cardoso}, {Carron}, {Chary}, {Chiang}, {Colombo}, {Combet}, {Crill},
  {Cuttaia}, {de Bernardis}, {de Zotti}, {Delabrouille}, {Delouis}, {Di
  Valentino}, {Dickinson}, {Diego}, {Dor{\'e}}, {Douspis}, {Ducout}, {Dupac},
  {Efstathiou}, {Elsner}, {En{\ss}lin}, {Eriksen}, {Falgarone}, {Fantaye},
  {Fernandez-Cobos}, {Ferri{\`e}re}, {Finelli}, {Forastieri}, {Frailis},
  {Fraisse}, {Franceschi}, {Frolov}, {Galeotta}, {Galli}, {Ganga},
  {G{\'e}nova-Santos}, {Gerbino}, {Ghosh}, {Gonz{\'a}lez-Nuevo}, {G{\'o}rski},
  {Gratton}, {Green}, {Gruppuso}, {Gudmundsson}, {Guillet}, {Handley},
  {Hansen}, {Helou}, {Herranz}, {Hivon}, {Huang}, {Jaffe}, {Jones},
  {Keih{\"a}nen}, {Keskitalo}, {Kiiveri}, {Kim}, {Krachmalnicoff}, {Kunz},
  {Kurki-Suonio}, {Lagache}, {Lamarre}, {Lasenby}, {Lattanzi}, {Lawrence}, {Le
  Jeune}, {Levrier}, {Liguori}, {Lilje}, {Lindholm}, {L{\'o}pez-Caniego},
  {Lubin}, {Ma}, {Mac{\'\i}as-P{\'e}rez}, {Maggio}, {Maino}, {Mandolesi},
  {Mangilli}, {Marcos-Caballero}, {Maris}, {Martin},
  {Mart{\'\i}nez-Gonz{\'a}lez}, {Matarrese}, {Mauri}, {McEwen}, {Melchiorri},
  {Mennella}, {Migliaccio}, {Miville-Desch{\^e}nes}, {Molinari}, {Moneti},
  {Montier}, {Morgante}, {Moss}, {Natoli}, {Pagano}, {Paoletti}, {Patanchon},
  {Perrotta}, {Pettorino}, {Piacentini}, {Polastri}, {Polenta}, {Puget},
  {Rachen}, {Reinecke}, {Remazeilles}, {Renzi}, {Ristorcelli}, {Rocha},
  {Rosset}, {Roudier}, {Rubi{\~n}o-Mart{\'\i}n}, {Ruiz-Granados}, {Salvati},
  {Sandri}, {Savelainen}, {Scott}, {Sirignano}, {Sunyaev}, {Suur-Uski},
  {Tauber}, {Tavagnacco}, {Tenti}, {Toffolatti}, {Tomasi}, {Trombetti},
  {Valiviita}, {Vansyngel}, {Van Tent}, {Vielva}, {Villa}, {Vittorio},
  {Wandelt}, {Wehus}, {Zacchei}, \& {Zonca}}]{2018arXiv180706212P}
{Planck Collaboration}, {Aghanim}, N., {Akrami}, Y., {et~al.}
  2018{\natexlab{b}}, arXiv e-prints, arXiv:1807.06212.
\newblock \doarXiv{1807.06212}

\bibitem[{{Predehl} \& {Schmitt}(1995)}]{1995A&A...293..889P}
{Predehl}, P., \& {Schmitt}, J.~H.~M.~M. 1995, \aap, 500, 459

\bibitem[{{Price} {et~al.}(2018){Price}, {Wurster}, {Tricco}, {Nixon},
  {Toupin}, {Pettitt}, {Chan}, {Mentiplay}, {Laibe}, {Glover}, {Dobbs},
  {Nealon}, {Liptai}, {Worpel}, {Bonnerot}, {Dipierro}, {Ballabio}, {Ragusa},
  {Federrath}, {Iaconi}, {Reichardt}, {Forgan}, {Hutchison}, {Constantino},
  {Ayliffe}, {Hirsh}, \& {Lodato}}]{2018PASA...35...31P}
{Price}, D.~J., {Wurster}, J., {Tricco}, T.~S., {et~al.} 2018, \pasa, 35, e031,
  \dodoi{10.1017/pasa.2018.25}

\bibitem[{{Reach} {et~al.}(2015){Reach}, {Heiles}, \&
  {Bernard}}]{2015ApJ...811..118R}
{Reach}, W.~T., {Heiles}, C., \& {Bernard}, J.-P. 2015, \apj, 811, 118,
  \dodoi{10.1088/0004-637X/811/2/118}

\bibitem[{{Redfield} {et~al.}(2004){Redfield}, {Wood}, \&
  {Linsky}}]{2004AdSpR..34...41R}
{Redfield}, S., {Wood}, B.~E., \& {Linsky}, J.~L. 2004, Advances in Space
  Research, 34, 41, \dodoi{10.1016/j.asr.2003.02.053}

\bibitem[{{Roman-Duval} {et~al.}(2010){Roman-Duval}, {Israel}, {Bolatto},
  {Hughes}, {Leroy}, {Meixner}, {Gordon}, {Madden}, {Paradis}, {Kawamura},
  {Li}, {Sauvage}, {Wong}, {Bernard}, {Engelbracht}, {Hony}, {Kim}, {Misselt},
  {Okumura}, {Ott}, {Panuzzo}, {Pineda}, {Reach}, \&
  {Rubio}}]{2010A&A...518L..74R}
{Roman-Duval}, J., {Israel}, F.~P., {Bolatto}, A., {et~al.} 2010, \aap, 518,
  L74, \dodoi{10.1051/0004-6361/201014575}

\bibitem[{{Savage} {et~al.}(1977){Savage}, {Bohlin}, {Drake}, \&
  {Budich}}]{1977ApJ...216..291S}
{Savage}, B.~D., {Bohlin}, R.~C., {Drake}, J.~F., \& {Budich}, W. 1977, \apj,
  216, 291, \dodoi{10.1086/155471}

\bibitem[{{Savage} \& {Sembach}(1991)}]{1991ApJ...379..245S}
{Savage}, B.~D., \& {Sembach}, K.~R. 1991, \apj, 379, 245,
  \dodoi{10.1086/170498}

\bibitem[{{Seligman} {et~al.}(2019){Seligman}, {Hopkins}, \&
  {Squire}}]{2019MNRAS.485.3991S}
{Seligman}, D., {Hopkins}, P.~F., \& {Squire}, J. 2019, \mnras, 485, 3991,
  \dodoi{10.1093/mnras/stz666}

\bibitem[{{Shingledecker} {et~al.}(2020){Shingledecker}, {Lamberts}, {Laas},
  {Vasyunin}, {Herbst}, {K{\"a}stner}, \& {Caselli}}]{2020ApJ...888...52S}
{Shingledecker}, C.~N., {Lamberts}, T., {Laas}, J.~C., {et~al.} 2020, \apj,
  888, 52, \dodoi{10.3847/1538-4357/ab5360}

\bibitem[{{Siebenmorgen} {et~al.}(2017){Siebenmorgen}, {Voschinnikov},
  {Bagnulo}, {Cox}, \& {Cami}}]{2017ques.workE..27S}
{Siebenmorgen}, R., {Voschinnikov}, N., {Bagnulo}, S., {Cox}, N., \& {Cami}, J.
  2017, in Submm/mm/cm QUESO Workshop 2017 (QUESO2017), 27,
  \dodoi{10.5281/zenodo.1038101}

\bibitem[{{Spitzer}(1954)}]{1954ApJ...120....1S}
{Spitzer}, Lyman, J. 1954, \apj, 120, 1, \dodoi{10.1086/145876}

\bibitem[{{Stone} {et~al.}(2008){Stone}, {Gardiner}, {Teuben}, {Hawley}, \&
  {Simon}}]{2008ApJS..178..137S}
{Stone}, J.~M., {Gardiner}, T.~A., {Teuben}, P., {Hawley}, J.~F., \& {Simon},
  J.~B. 2008, \apjs, 178, 137, \dodoi{10.1086/588755}

\bibitem[{{Thoraval} {et~al.}(1997){Thoraval}, {Boisse}, \&
  {Duvert}}]{1997A&A...319..948T}
{Thoraval}, S., {Boisse}, P., \& {Duvert}, G. 1997, \aap, 319, 948

\bibitem[{{Thoraval} {et~al.}(1999){Thoraval}, {Boiss{\'e}}, \&
  {Duvert}}]{1999A&A...351.1051T}
{Thoraval}, S., {Boiss{\'e}}, P., \& {Duvert}, G. 1999, \aap, 351, 1051

\bibitem[{van~der Walt {et~al.}(2014)van~der Walt, {S}ch\"onberger,
  {Nunez-Iglesias}, {B}oulogne, {W}arner, {Y}ager, {G}ouillart, {Y}u, \& the
  scikit-image contributors}]{scikit-image}
van~der Walt, S., {S}ch\"onberger, J.~L., {Nunez-Iglesias}, J., {et~al.} 2014,
  PeerJ, 2, e453, \dodoi{10.7717/peerj.453}

\bibitem[{{Wardle}(2007)}]{2007Ap&SS.311...35W}
{Wardle}, M. 2007, \apss, 311, 35, \dodoi{10.1007/s10509-007-9575-8}

\bibitem[{{Weidenschilling}(1977)}]{1977MNRAS.180...57W}
{Weidenschilling}, S.~J. 1977, \mnras, 180, 57, \dodoi{10.1093/mnras/180.1.57}

\bibitem[{{Weingartner} \& {Draine}(2001)}]{2001ApJS..134..263W}
{Weingartner}, J.~C., \& {Draine}, B.~T. 2001, \apjs, 134, 263,
  \dodoi{10.1086/320852}

\bibitem[{{Welsh} {et~al.}(2010){Welsh}, {Lallement}, {Vergely}, \&
  {Raimond}}]{2010A&A...510A..54W}
{Welsh}, B.~Y., {Lallement}, R., {Vergely}, J.~L., \& {Raimond}, S. 2010, \aap,
  510, A54, \dodoi{10.1051/0004-6361/200913202}

\bibitem[{{Wolfire} {et~al.}(1995){Wolfire}, {Hollenbach}, {McKee}, {Tielens},
  \& {Bakes}}]{1995ApJ...443..152W}
{Wolfire}, M.~G., {Hollenbach}, D., {McKee}, C.~F., {Tielens}, A. G. G.~M., \&
  {Bakes}, E. L.~O. 1995, \apj, 443, 152, \dodoi{10.1086/175510}

\end{thebibliography}
\bibliographystyle{aasjournal}



\end{document}